\documentclass{article}

\usepackage[utf8]{inputenc}
\usepackage[T1]{fontenc}
\usepackage{microtype}
\usepackage{graphicx}
\usepackage{booktabs}
\usepackage{multirow}
\usepackage{amsmath,amssymb,amsthm,mathtools}
\usepackage{bm}
\usepackage{xcolor}
\usepackage[colorlinks=true,citecolor=blue!60!black,linkcolor=blue!60!black,urlcolor=blue!60!black]{hyperref}
\usepackage{icml_style}

\theoremstyle{plain}
\newtheorem{theorem}{Theorem}
\newtheorem{proposition}[theorem]{Proposition}

\theoremstyle{definition}

\theoremstyle{remark}
\newtheorem{remark}[theorem]{Remark}

\newcommand{\R}{\mathbb{R}}
\newcommand{\cube}{[0,1]^d}
\newcommand{\clip}{\operatorname{clip}_{[0,1]}}
\newcommand{\Restore}{\operatorname{Restore}}
\newcommand{\Id}{\mathrm{Id}}
\newcommand{\E}{\mathbb{E}}

\icmltitlerunning{Equivariant Geometric Graph Operators for Incremental and Generalizable Minimax Design}

\begin{document}

\icmltitle{Equivariant Geometric Graph Operators for Incremental and\\ Generalizable Minimax Space-Filling Design}

\begin{abstract}
Minimax distance designs minimize the covering radius (fill distance) of an $n$-point configuration in $[0,1]^d$, providing worst-case error guarantees for surrogate modeling and computer experiments. Computing such designs across varying sample budgets $n$ and dimensions $d$ remains challenging because the objective is a non-smooth, non-convex min-max-min problem; classical solvers incur high per-instance cost or restrict $(n,d)$ to lattice sizes, while existing neural point-set samplers typically fix $(n,d)$ and optimize discrepancy rather than fill distance. We present a unified deep learning framework for amortized and incremental minimax design built on three components: (1) a lightweight, design-only ($X$-only) geometric graph neural operator (\textbf{V7}) whose layers are strictly $(S_n \times B_d)$-equivariant to point permutations and all $2^d d!$ hypercube symmetries with parameter count independent of $(n,d)$, paired with finite-step hard feasibility restoration and a $C^1$-smooth local log-barrier; (2) \emph{incremental refinement} from state-of-the-art classical designs via a zero-initialized residual gate and a shared-witness asymmetric non-degradation objective; and (3) \emph{generalizability} across arbitrary initializations via shape-matched structured/random training and multi-loop iterative self-feedback ($X^{(k)} = R_\theta(X^{(k-1)})$) with witness-guarded selection. Across 2,041 held-out specifications ($d \in [2,12], n \in [10,300]$), a 1,130-parameter V7 operator outperforms six larger bipartite and Transformer baselines on 94.66\% of specifications (and Sobol' on 96.52\%), including zero-shot unseen dimensions $d \in \{6,11,12\}$. Under incremental refinement, V7 improves Mak--Joseph clustering+PSO designs on 8/9 held-out specifications ($-2.62\%$, $p=0.012$) while increasing pairwise separation. Crucially, combining random-input mixing with multi-loop self-feedback enables a 14.4K-parameter V7 operator to outperform Mak--Joseph clustering+PSO from \emph{pure uniform random noise} on 6/9 held-out specifications in two open-loop passes ($+1.69\%$) and 7/9 under witness-guarded selection ($+3.34\%$), running $530\times$--$11{,}900\times$ faster on GPU.
\end{abstract}

\section{Introduction}
\label{sec:intro}

One-shot space-filling experimental designs select an entire batch of $n$ input points $X = \{x_1, \dots, x_n\} \subset \cube$ prior to evaluating an expensive computer simulator \citep{santner2018design,huan2024optimal}. Among space-filling criteria, \emph{minimax distance design} \citep{johnson1990minimax} minimizes the Euclidean fill distance (covering radius)
\begin{equation}
\label{eq:fill_def}
h(X) \;=\; \sup_{u \in \cube} \min_{i \in \{1,\dots,n\}} \|u - x_i\|_2.
\end{equation}
Fill distance directly controls worst-case interpolation error for Gaussian process surrogates and reproducing kernel Hilbert spaces \citep{johnson1990minimax,wendland2004scattered}: reducing $h(X)$ shrinks the largest uncovered hole in the design domain.

Despite its theoretical centrality, computing high-quality minimax designs for arbitrary $(n,d)$ is notoriously difficult. Evaluating $h(X)$ itself requires solving a global maximization over $\cube$, turning minimax design into a non-smooth, non-convex min-max-min bilevel problem. Classical solvers fall into three regimes: (i) fast constructive or low-discrepancy baselines such as Latin hypercube sampling \citep{mckay1979comparison}, scrambled Sobol' sequences \citep{joe2008sobol}, and farthest-first $k$-center heuristics \citep{gonzalez1985clustering,shang2021fssf}, which scale well but leave large covering radii; (ii) per-instance iterative optimizers such as clustering with particle swarm optimization (PSO) \citep{mak2018minimax}, which yield strong designs for arbitrary $(n,d)$ at the cost of tens of seconds per specification; and (iii) algebraic interleaved lattice constructions \citep{he2017interleaved,he2024interleaved}, which achieve near-optimal covering radii when $d \le 8$ and $n$ matches admissible lattice counts, but cannot target arbitrary $(n,d)$ without padding or truncation.

\begin{figure*}[t]
\centering
\includegraphics[width=\textwidth]{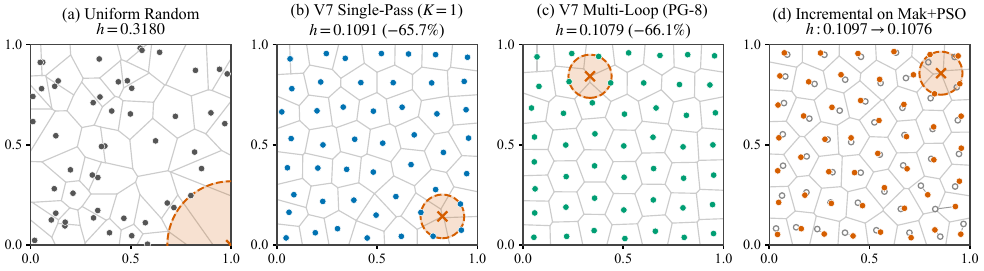}
\caption{\textbf{Real 2D Voronoi tessellations and exact covering radii on held-out $(d=2, n=50)$} produced by our 14.4K-parameter V7 operator ($16\times 128$, 10h mixed training). Dashed orange circles mark the exact largest empty ball (true fill distance $h(X)$, center marked $\times$). \textbf{(a)} Uniform random input ($h=0.3180$). \textbf{(b)} Single forward pass $K=1$ ($h=0.1091$, $-65.7\%$, already surpassing Mak+PSO's $0.1097$ in $7.2\,\mathrm{ms}$). \textbf{(c)} Witness-guarded multi-loop PG-8 ($h=0.1079$, $-66.1\%$). \textbf{(d)} Single-pass incremental refinement directly on a Mak+PSO design (hollow circles $\to$ filled dots, reducing $h$ from $0.1097$ to $0.1076$).}
\label{fig:voronoi}
\end{figure*}

Recently, deep learning has emerged for one-shot point-set generation, notably Message-Passing Monte Carlo (MPMC) \citep{rusch2024mpmc} and Neural Low-Discrepancy Sequences (NeuroLDS) \citep{vanhuffel2026neurolds}. However, these methods optimize differentiable \emph{discrepancy} criteria rather than minimax fill distance, and train separate models for fixed $(n,d)$ or fixed dimension and maximum sequence length. Learning a \emph{single} neural operator that (a) generalizes zero-shot across unseen budgets $n$ and dimensions $d$, (b) strictly improves strong classical designs rather than discarding them, and (c) generates competitive minimax designs even from arbitrary random noise remains an open methodological challenge.

In this paper, we build a unified deep learning methodology for minimax distance design structured around three synergistic pillars (illustrated in Figure~\ref{fig:voronoi}):

\begin{enumerate}
    \item \textbf{Symmetry-Equivariant $X$-Only Geometric Graph Operator with Exact Feasibility (V7).} Comparing seven architectural iterations (V1--V7), we find that decoupling adversarial witness search (training only) from a pure design-to-design geometric GNN parameterized in the natural length scale $s(n,d) = n^{-1/d}$ yields dramatic gains in accuracy and parameter efficiency. We prove that V7 is strictly $(S_n \times B_d)$-equivariant to all $n!$ point permutations and all $2^d d!$ hypercube symmetries with parameter count $L(7W+2)$ independent of $(n,d)$ (Theorem~\ref{thm:equivariance}), whereas standard coordinate-encoding Transformers break reflection symmetry $\mathbb{Z}_2^d$. To eliminate exact point collisions without distorting well-separated configurations, we combine a finite-step dyadic feasibility restoration map with a $C^1$-smooth local squared log-barrier (Theorem~\ref{thm:feasibility}).
    \item \textbf{Incremental Learning from Best Existing Designs.} Rather than competing with classical DOE solvers from scratch, we wrap V7 in a zero-initialized residual gate ($R_{\theta_0} = \Id$) and train it directly on strong classical anchors (Mak--Joseph clustering+PSO, He's interleaved lattices, and scrambled Sobol') using a \emph{shared-witness asymmetric non-degradation objective}. We prove that evaluating input and output on the union witness set $U(X,Z) = \mathcal{S}(X)\cup\mathcal{S}(Z)$ eliminates witness-mismatch variance at the identity anchor and enforces monotone coverage improvement (Theorem~\ref{thm:shared_witness}).
    \item \textbf{Generalizability via Shape-Matched Random Mixing and Multi-Loop Self-Feedback.} Training solely on high-quality structured anchors causes severe covariate shift on random inputs ($-31.7\%$ vs.\ Mak+PSO). Pairing every structured batch with a shape-matched uniform random batch yields a $+34.9$ percentage-point leap on random inputs while also boosting structured refinement. Because the resulting operator maps across the entire quality spectrum, feeding its output recursively back as input—\emph{multi-loop self-feedback}, $X^{(k)} = R_\theta(X^{(k-1)})$—progressively lowers $h(X)$. We characterize its two-regime contraction and prove an exact performance bound for \emph{witness-guarded} multi-loop selection (Proposition~\ref{prop:multiloop}).
    \item \textbf{Extensive Cross-Dimension, Incremental, and Multi-Loop Validation.} Across 3,201 fixed-benchmark specifications ($2{,}041$ held-out, including zero-shot unseen dimensions $d \in \{6,11,12\}$), controlled 5-arm ablations, a 19-trial Bayesian architecture search, and a 10-hour compute-matched comparison against Transformers, V7 consistently dominates attention baselines with $85\times$--$5{,}780\times$ fewer parameters. On nine held-out even specifications starting from pure uniform random noise, a 14.4K-parameter V7 operator beats Mak+PSO on 6/9 specs in two open-loop passes ($+1.69\%$) and 7/9 under witness-guarded selection ($+3.34\%$), at $530\times$--$11{,}960\times$ lower GPU latency.
\end{enumerate}

\section{Problem Formulation and Related Work}
\label{sec:related}

\paragraph{Minimax Distance Design.}
Given budget $n \ge 2$ and dimension $d \ge 2$, a continuous one-shot design is a matrix $X \in \bigl(\cube\bigr)^n \equiv [0,1]^{n \times d}$ whose rows $x_1, \dots, x_n \in \cube$ represent unordered experimental runs. Our primary criterion is the Euclidean fill distance $h(X)$ in \eqref{eq:fill_def}. Because minimizing $h(X)$ alone can leave pairs of points nearly coincident—wasting experimental budget and inducing numerical ill-conditioning in surrogate kernel matrices—we jointly track the \emph{separation distance} (minimum pairwise distance) and \emph{mesh ratio}:
\begin{equation}
\label{eq:sep_mesh}
\delta(X) = \min_{1 \le i < j \le n} \|x_i - x_j\|_2, \qquad \rho(X) = \frac{2 h(X)}{\delta(X)}.
\end{equation}
A feasible design for deterministic computer experiments must satisfy $\delta(X) > 0$ (no duplicate runs). Notice that both $h(X)$ and $\delta(X)$ are invariant under the product group $S_n \times B_d$, where $S_n$ permutes rows and the hyperoctahedral group $B_d = (\mathbb{Z}_2)^d \rtimes S_d$ permutes columns and reflects any subset of coordinates via $x_{i,m} \mapsto 1 - x_{i,m}$.

\paragraph{Unified Evaluation Protocol.}
In $d=2$, we compute $h(X)$ \emph{exactly} via bounded Voronoi tessellation over $[0,1]^2$ (checking all Voronoi vertices in $[0,1]^2$, bisector-boundary intersections, and the four corners). In $d \ge 3$, exact Voronoi computation is intractable at scale; following a fixed deterministic protocol across all methods, we compute a tight lower bound using $4{,}096$ scrambled Sobol' interior points, $1{,}024$ face projections, and all $2^d$ hypercube corners (or $4{,}096$ sampled corners for $d>12$), followed by 12 steps of projected distance ascent from the 32 farthest candidates with step schedule $0.12\,n^{-1/d}\cdot 0.85^k$. Separation $\delta(X)$ is computed exactly via KD-tree nearest-neighbor queries in FP64.

\paragraph{Classical Minimax and Space-Filling Methods.}
Foundational space-filling literature distinguishes minimax (covering) from maximin (separation) designs \citep{johnson1990minimax,morris1995exploratory} and projection criteria such as MaxPro \citep{joseph2015maxpro} and uniform discrepancy \citep{fang2000uniform}. For minimax distance specifically, \citet{mak2018minimax} proposed clustering-based $C_q$-center optimization hybridized with PSO (and \texttt{SFDesign} \citealt{wang2026sfdesign}); \citet{shang2021fssf} introduced fully sequential forward-selection (\texttt{FSSF}) with boundary reflection; \citet{he2017interleaved,he2024interleaved} constructed interleaved lattices with provably strong covering radii and fill-separation trade-offs for $d \le 8$; and \citet{du1999centroidal} studied centroidal Voronoi tessellations (Lloyd's method). Unlike classical solvers that optimize each $(n,d)$ independently or restrict $n$ to lattice grids, our learned operator amortizes minimax optimization across $(n,d)$ and can refine classical designs directly.

\paragraph{Deep Learning for Experimental Design and Point Sets.}
Direct neural point-set generators include Deep Point Correlation Design \citep{leimkuhler2019dpcd} for periodic spectral textures, MPMC \citep{rusch2024mpmc} using GNNs trained on differentiable $L_2$-discrepancy, NeuroLDS \citep{vanhuffel2026neurolds} mapping indices to finite low-discrepancy prefixes, diffusion-based 2D sampling \citep{doignies2023example}, differentiable Owen scrambling \citep{doignies2024differentiable}, gradient post-processing \citep{clement2025postprocessing}, permutation search \citep{clement2025searching}, and LLM program synthesis \citep{sadikov2026llm}. In statistical optimal design, NBDO \citep{ladas2026nbdo} performs latent-space BO for $A$-optimal designs, PIED \citep{hemachandra2025pied} optimizes one-shot PDE sensors, and amortized sequential Bayesian experimental design \citep{foster2021dad,hedman2025stepdad,saravanan2026diffbed} learns adaptive policies conditioned on observed responses. Our work differs fundamentally by targeting \emph{continuous minimax covering} with a single $(S_n \times B_d)$-equivariant operator that generalizes zero-shot across both $n$ and $d$, supports incremental warm-start refinement, and composes iteratively.

\paragraph{Equivariant Set Architectures and Learned Optimizers.}
Permutation-equivariant networks such as Deep Sets \citep{zaheer2017deepsets}, Set Transformer \citep{lee2019settransformer}, and $E(n)$-Equivariant GNNs (EGNN) \citep{satorras2021egnn} provide strong geometric inductive biases. Our V7 operator adapts radial-displacement message passing specifically to the \emph{bounded hypercube} $\cube$ by incorporating scale-normalized $n^{-1/d}$ distances and an odd boundary-repulsion field preserving $B_d$. Finally, our multi-loop operator connects to learned optimizers \citep{andrychowicz2016l2l} and neural local improvement \citep{chen2019neurewriter,kwon2020pomo}, extending iterative neural refinement to continuous min-max-min geometry.

\section{Methodology}
\label{sec:method}

We now detail the three pillars of our framework: (i) the core $(S_n \times B_d)$-equivariant V7 geometric graph operator and feasibility mechanisms, (ii) residual-gated incremental learning from classical designs, and (iii) generalizability via shape-matched random mixing and multi-loop self-feedback.

\subsection{Core Architecture: Decoupled $X$-Only Geometric GNN (V7)}
\label{sec:v7}

A natural first instinct for minimax design is a bipartite design--probe network that jointly updates design points $X$ and adversarial probe points $U$ during inference (our V1--V6 baselines; Appendix~\ref{app:archs}). Empirically, however, co-training probe heads adds optimization instability and requires expensive probe search or extra parameters at inference time. Our key architectural insight in \textbf{V7} is to \emph{completely decouple adversarial witness search (used strictly during training) from a lightweight, pure design-to-design ($X$-only) geometric message-passing operator}.

Let $s = s(n,d) = n^{-1/d}$ denote the characteristic inter-point spacing of $n$ points in $\cube$. Given an input design $X^{(0)} \in \bigl(\cube\bigr)^n$, V7 applies $L$ geometric message-passing layers. At layer $\ell \in \{1,\dots,L\}$ with current coordinates $X = X^{(\ell-1)}$, we compute pairwise scaled displacements, squared radii, Gaussian locality weights, local densities, and scaled nearest-face distances:
\begin{align}
\Delta_{ij} &= \frac{x_i - x_j}{s} \in \R^d, \quad r_{ij}^2 = \|\Delta_{ij}\|_2^2, \quad k_{ij} = e^{-r_{ij}^2}, \label{eq:v7_geom}\\
\rho_i &= \sum_{j=1}^n k_{ij}, \qquad w_i = \frac{1}{s}\min_{m \in \{1,\dots,d\}} \min(x_{i,m}, 1 - x_{i,m}). \label{eq:v7_node}
\end{align}
Each directed edge $(i,j)$ forms a 5-dimensional \emph{dimension-independent} invariant feature vector
\begin{equation}
\label{eq:v7_edge}
e_{ij} = \big(\min(r_{ij}^2, 12),\; \rho_i,\; \rho_j,\; w_i,\; w_j\big) \in \R^5.
\end{equation}
A layer-specific scalar edge MLP $\phi_\ell : \R^5 \to \R$ (`Linear(5, W)` $\to$ `SiLU` $\to$ `Linear(W, 1)`) modulates the locality kernel to form an equivariant pairwise interaction vector, combined with a learnable coordinate-wise boundary field parameterized by a single scalar $\gamma_\ell \in \R$:
\begin{align}
v_i &= \sum_{j=1}^n k_{ij}\,\phi_\ell(e_{ij})\,\Delta_{ij}, \label{eq:v7_msg}\\
b_{i,m} &= e^{-2 x_{i,m}/s} - e^{-2(1 - x_{i,m})/s}, \quad m=1,\dots,d, \label{eq:v7_wall}\\
x_i^{(\ell)} &= \clip\!\Big(x_i^{(\ell-1)} + s\tanh\!\big(v_i + \gamma_\ell b_i\big)\Big). \label{eq:v7_upd}
\end{align}
Denote the $L$-layer map by $G_\theta(X^{(0)}) = X^{(L)}$. Every layer executes one all-pairs geometric interaction in $\mathcal{O}(n^2(d+W))$ time without invoking any probe network at inference.

\paragraph{Decoupled Adversarial Witness Search (Training Only).}
During base V7 training on output $Z = G_\theta(X^{(0)})$, we detach $Z$ from the computation graph and run a lightweight projected witness ascent $\mathcal{S}(Z)$: sample $M_{\mathrm{pool}} = 512$ mixed interior/face/corner candidates, select the $M_w = 48$ points farthest from $Z$, and perform $T_w = 8$ steps of normalized nearest-site distance ascent $u \leftarrow \clip(u + 0.32\,s\cdot 0.78^t \frac{u - z_{\mathrm{nn}}(u)}{\|u - z_{\mathrm{nn}}(u)\|_2})$, retaining the best iterate per start. Fixing the resulting witness set $U = \mathcal{S}(Z)$, base V7 minimizes the scale-normalized smooth maximum squared distance:
\begin{equation}
\label{eq:base_loss}
\mathcal{L}_{\mathrm{cov}}(Z; U) = \frac{1}{\beta}\log \sum_{u \in U} \exp\!\Big(\beta\,\frac{n^{2/d}}{d}\min_{1 \le i \le n}\|u - z_i\|_2^2\Big),
\end{equation}
with inverse temperature $\beta = 5$ (and exact maximum $\max_{u \in U}$ in the paired incremental loss below). The factor $n^{2/d}/d$ normalizes squared covering radii across arbitrary $(n,d)$, since $\E[\min_i \|u-x_i\|_2^2] = \Theta(d\,n^{-2/d})$.

\begin{theorem}[Exact $(S_n \times B_d)$-Equivariance and Size/Dimension Independence]
\label{thm:equivariance}
Let $S_n$ act on $X \in \bigl(\cube\bigr)^n$ by row permutations and let $B_d = (\mathbb{Z}_2)^d \rtimes S_d$ act by column permutations and coordinate reflections $x_{i,m} \mapsto 1 - x_{i,m}$. Then in exact real arithmetic:
\begin{enumerate}
    \item[(i)] Both $h(X)$ and $\delta(X)$ are strictly $(S_n \times B_d)$-invariant.
    \item[(ii)] Every V7 layer \eqref{eq:v7_geom}--\eqref{eq:v7_upd}, the generator $G_\theta$, the hard feasibility wrapper $H_\theta$ \eqref{eq:restore}, and the residual-gated refiner $R_\theta$ \eqref{eq:mak_refiner} are strictly $(S_n \times B_d)$-equivariant: $R_\theta(P \cdot Q(X)) = P \cdot Q(R_\theta(X))$ for all $(P, Q) \in S_n \times B_d$.
    \item[(iii)] The parameter count of $G_\theta$ is $|\theta_G| = L(7W + 2)$ (and $|\theta_R| = L(7W+2)+1$), strictly independent of both $n$ and $d$.
\end{enumerate}
\end{theorem}
\emph{Proof sketch.} Pairwise squared distances $r_{ij}^2$, densities $\rho_i$, and boundary distances $w_i$ are invariant to $B_d$, so scalar weights $k_{ij}\phi_\ell(e_{ij})$ are $B_d$-invariant and $S_n$-equivariant. Relative displacements $\Delta_{ij}$ and the odd wall field $b_i$ transform equivariantly under axis permutations and sign flips; since $\tanh$ is odd and $\clip(1-y) = 1-\clip(y)$, each layer preserves $S_n \times B_d$. Full proof in Appendix~\ref{app:proofs}.

\begin{remark}[Why Coordinate-Encoding Transformers Fail Reflection Equivariance]
\label{rem:transf_sym}
Dimension-agnostic Transformers such as \texttt{TransformerRefiner} and \texttt{GeometricTransformer v2} (Section~\ref{sec:exp_mix}) encode each coordinate via a shared MLP applied to $(x_{i,m}, \min(x_{i,m}, 1-x_{i,m}))$. Because raw $x_{i,m}$ enters nonlinear layers directly, the encoder breaks $\mathbb{Z}_2^d$ reflection equivariance, forcing the network to learn across $2^d$ octant reflections (e.g., $1{,}024$ in $d=10$, $4{,}096$ in $d=12$). V7 avoids raw coordinates entirely, using only relative displacements $\Delta_{ij}$, reflection-invariant $w_i$, and the odd boundary force $b_{i,m}$.
\end{remark}

\subsection{Hard Feasibility Recovery and Local Log-Barrier}
\label{sec:feasibility}

Unconstrained coverage minimization occasionally merges two points onto identical coordinates ($\delta(Z) = 0$): in our fixed benchmark, original V7 exhibits exact collisions on 470/2,041 held-out layouts despite achieving the lowest fill distance (Section~\ref{sec:exp_fixed}). Simply clipping or adding a smooth global inverse-distance penalty $1/(\epsilon + \delta/b_{n,d})$ reduces but does not eliminate collisions (163/2,041 remain) and introduces global repulsive bias everywhere.

We solve this cleanly with two coupled mechanisms:

\paragraph{1. Dyadic Hard Feasibility Restoration.}
Given any proposal $Z \in \bigl(\cube\bigr)^n$ and a distinct input anchor $X^{(0)} \in \bigl(\cube\bigr)^n$ ($\delta(X^{(0)}) > 0$), we test for exact row duplicates in working precision. If any collision occurs, we backtrack along the convex segment toward $X^{(0)}$ over the dyadic grid $\mathcal{A} = \{2^{-5}, 2^{-4}, 2^{-3}, 2^{-2}, 2^{-1}, 1\}$:
\begin{equation}
\label{eq:restore}
\Restore(Z, X^{(0)}) = (1 - \alpha^*) Z + \alpha^* X^{(0)},
\end{equation}
where $\alpha^* = \min\{\alpha \in \mathcal{A} : \delta((1-\alpha)Z + \alpha X^{(0)}) > 0\}$. We denote $H_\theta(X^{(0)}) = \Restore(G_\theta(X^{(0)}), X^{(0)})$. During training, gradients backpropagate directly through the restored feasible coordinates (with straight-through clipping on the separation path and unit input-direction subgradients $\frac{x_i^{(0)} - x_j^{(0)}}{\|x_i^{(0)} - x_j^{(0)}\|_2}$ if a pre-restoration pair coincides).

\paragraph{2. Calibrated $C^1$ Local Squared Log-Barrier.}
To discourage \emph{near}-collisions without perturbing well-separated minimax placements, we calibrate a Monte Carlo reference scale $b_{n,d}$ as the median minimum pairwise distance over 256 i.i.d.\ uniform designs in $\cube$ (precomputed once for $d \in [2,12], n \in [10,300]$) and penalize only pairs below a dimensionless cutoff $\tau = 0.5$:
\begin{equation}
\label{eq:log_barrier}
\psi_{\mathrm{log}}\!\Big(\frac{\delta(Z)}{b_{n,d}}\Big) = \left(\log \operatorname{clamp}\!\Big(\frac{\delta(Z)}{\tau\, b_{n,d}},\; 10^{-4},\; 1\Big)\right)^2.
\end{equation}

\begin{theorem}[Finite-Step Feasibility and Zero-Bias Barrier]
\label{thm:feasibility}
For any distinct anchor $X^{(0)} \in \bigl(\cube\bigr)^n$ ($\delta(X^{(0)}) > 0$) and any proposal $Z \in \bigl(\cube\bigr)^n$:
\begin{enumerate}
    \item[(i)] Every iterate $Z(\alpha) = (1-\alpha)Z + \alpha X^{(0)}$ lies in $\bigl(\cube\bigr)^n$, and $\Restore(Z, X^{(0)})$ terminates in at most $|\mathcal{A}| = 6$ steps with $\delta(\Restore(Z, X^{(0)})) > 0$. Furthermore, over $\R$, at most $\binom{n}{2}$ values of $\alpha \in [0,1)$ can produce a collision.
    \item[(ii)] Ignoring the numerical floor $10^{-4}$, the unclamped local barrier $\tilde{\psi}_{\mathrm{log}}(r) = (\log \min(r/\tau, 1))^2$ is $C^1$-smooth on $(0,\infty)$, satisfies $\lim_{r\to 0^+}\tilde{\psi}_{\mathrm{log}}(r) = +\infty$, and achieves exact zero value and zero derivative ($\tilde{\psi}_{\mathrm{log}}(r) = \tilde{\psi}'_{\mathrm{log}}(r) = 0$) for all $r \ge \tau$.
\end{enumerate}
\end{theorem}
Theorem~\ref{thm:feasibility}(ii) explains why replacing the global reciprocal penalty $1/(\epsilon + r)$ (whose derivative $-1/(\epsilon + r)^2 < 0$ biases all pairs at all distances) with $\psi_{\mathrm{log}}$ simultaneously improves \emph{both} fill distance and near-collision counts in paired experiments (Table~\ref{tab:fixed_sep}).

\subsection{Incremental Learning from Best Existing Designs}
\label{sec:incremental}

Classical minimax solvers such as Mak--Joseph clustering+PSO \citep{mak2018minimax} and He's interleaved lattices \citep{he2017interleaved} already produce high-quality designs on many specifications. Rather than discarding this domain knowledge, we formulate \emph{amortized incremental refinement}: learn a single operator that accepts any strong classical design $X$ and reliably improves its covering radius in one millisecond forward pass.

\paragraph{Zero-Initialized Residual-Gated Operator.}
Given a base Hard V7 backbone $H_\theta$, we introduce a learnable scalar gate $g \in \R$ initialized at $g = 0$ (or $g = 0.1$ when training from scratch):
\begin{equation}
\label{eq:mak_refiner}
R_\theta(X) = \Restore\!\Big(\clip\!\big(X + \tanh(g)\,(H_\theta(X) - X)\big),\; X\Big).
\end{equation}
At initialization $g = 0$ (warm-started from a pre-trained V7 backbone), $R_{\theta_0}(X) \equiv X$ identically for all $X \in \bigl(\cube\bigr)^n$. Thus training begins strictly at the performance level of the best existing classical designs with zero initial regression.

\paragraph{Shared-Witness Asymmetric Non-Degradation Objective.}
When refining an already-strong anchor $X$, optimizing $Z = R_\theta(X)$ against freshly sampled witnesses $\mathcal{S}(Z)$ alone can degrade coverage on holes that $X$ previously covered well. We therefore construct the \emph{union witness set} $U(X, Z) = \mathcal{S}(X) \cup \mathcal{S}(\operatorname{sg}(Z))$ (where $\operatorname{sg}$ stops gradients), and define the normalized empirical squared fill distance on $U$:
\begin{equation}
\label{eq:emp_fill}
\hat{h}_U^2(Y) = \frac{n^{2/d}}{d} \max_{u \in U} \min_{1 \le i \le n} \|u - y_i\|_2^2.
\end{equation}
We train $R_\theta$ by minimizing the asymmetric paired objective (with $\lambda_{\mathrm{reg}} = 4.0$, $\lambda_{\mathrm{sep}} = 0.02$):
\begin{equation}
\label{eq:inc_loss}
\mathcal{L}_{\mathrm{inc}}(\theta; X) = \hat{h}_{U(X,Z)}^2(Z) + \lambda_{\mathrm{reg}}\,\big[\hat{h}_{U(X,Z)}^2(Z) - \hat{h}_{U(X,Z)}^2(X)\big]_+ + \lambda_{\mathrm{sep}}\,\psi_{\mathrm{log}}\!\Big(\frac{\delta(Z)}{b_{n,d}}\Big),
\end{equation}
where $[\cdot]_+ = \max(\cdot, 0)$, optimized via AdamW with FP64 global gradient-norm clipping (threshold 2.0) to prevent FP32 norm overflow on large $(n,d)$ batches.

\begin{theorem}[Shared-Witness Paired Surrogate Properties]
\label{thm:shared_witness}
Let $U = U(X, Z) = \mathcal{S}(X) \cup \mathcal{S}(Z)$ and $\hat{h}_U(Y) = \max_{u \in U}\min_i \|u - y_i\|_2$. Then:
\begin{enumerate}
    \item[(i)] \textbf{Tighter Lower Bound:} $\hat{h}_U(Z) \ge \max\{\hat{h}_{\mathcal{S}(X)}(Z), \hat{h}_{\mathcal{S}(Z)}(Z)\}$ and $\hat{h}_U(Y) \le h(Y)$ for $Y \in \{X, Z\}$, simultaneously penalizing unresolved holes of $X$ and newly created holes of $Z$.
    \item[(ii)] \textbf{Zero Variance at the Identity Anchor:} At $Z = X$ (e.g., $g = 0$), the paired margin satisfies $\hat{h}_U^2(Z) - \hat{h}_U^2(X) \equiv 0$ identically for every random realization of $U$, whereas independent witness sets $U_1, U_2$ incur non-zero random bias $\E\big[[\hat{h}_{U_2}^2(X) - \hat{h}_{U_1}^2(X)]_+\big] > 0$.
    \item[(iii)] \textbf{Asymmetric Descent Coupling:} Whenever $\hat{h}_U^2(Z) > \hat{h}_U^2(X)$, the subgradient with respect to $Z$ is scaled by $(1 + \lambda_{\mathrm{reg}}) = 5\times$ specifically along the active worst-case witness $u^* \in \arg\max_{u\in U}\min_i \|u-z_i\|_2$, focusing capacity on eliminating coverage regressions.
\end{enumerate}
\end{theorem}

\subsection{Generalizability: Random Mixing and Multi-Loop Self-Feedback}
\label{sec:generalizability}

While incremental training on structured anchors (Mak+PSO and He lattices) yields strong local refinement ($+6.4\%$ over Mak+PSO), we observe a striking failure mode (Section~\ref{sec:exp_mix}): an operator trained \emph{exclusively} on high-quality anchors suffers severe covariate shift when deployed on uniform random initializations, lagging Mak+PSO by $17.6\%$--$31.7\%$. To build a truly \emph{generalizable} minimax design operator that excels from \emph{both} best-known classical designs and arbitrary random inputs, we introduce two complementary mechanisms.

\paragraph{1. Shape-Matched Structured/Random Input Mixing.}
At each training iteration, we pair a structured anchor mini-batch $X_{\mathrm{struct}} \in [0,1]^{B \times n \times d}$—cycling across Mak+PSO, He (2017) interleaved lattices, and scrambled Sobol' designs—with an identically shaped uniform random mini-batch $X_{\mathrm{rand}} \sim \mathrm{Uniform}([0,1])^{B \times n \times d}$, optimizing both under the exact same paired-witness objective \eqref{eq:inc_loss}. To maximize GPU throughput across variable $n \in [10, 300]$ without memory spikes, we dynamically set batch size via an edge-budget controller $B = \operatorname{clamp}(\lfloor E / n^2 \rfloor, 1, B_{\max})$ (with automatic recursive halving on OOM). Crucially, exposing the equivariant GNN to both endpoints of the design-quality spectrum—coarse random clouds and near-optimal structured lattices—teaches a unified vector field that performs large global hole-filling moves when inputs are irregular and fine local equilibration when inputs are already well-spaced.

\begin{algorithm}[t]
\caption{Training and Multi-Loop Inference for Equivariant Minimax Operator ($R_\theta$)}
\label{alg:framework}
\begin{algorithmic}[1]
\STATE \textbf{Training (Incremental + Shape-Matched Mixing):}
\STATE Initialize $R_\theta$ \eqref{eq:mak_refiner} from V7 backbone ($g=0$) or scratch ($g=0.1$).
\FOR{step $t = 1, 2, \dots$}
    \STATE Cycle structured source $s \in \{\text{Mak}, \text{He}, \text{Sobol'}\}$; sample $(d,n)$ from odd-$n$ train split and set $B \leftarrow \operatorname{clamp}(\lfloor E/n^2\rfloor, 1, B_{\max})$.
    \STATE Draw $X_{\mathrm{struct}} \in [0,1]^{B\times n\times d}$ and $X_{\mathrm{rand}} \sim \mathrm{Unif}(0,1)^{B\times n\times d}$.
    \FOR{$X \in \{X_{\mathrm{struct}}, X_{\mathrm{rand}}\}$}
        \STATE Forward: $Z \leftarrow R_\theta(X)$ via V7 layers \eqref{eq:v7_geom}--\eqref{eq:v7_upd} and $\Restore$ \eqref{eq:restore}.
        \STATE Search union witnesses $U \leftarrow \mathcal{S}(X) \cup \mathcal{S}(\operatorname{sg}(Z))$ (no grad).
        \STATE Compute $\mathcal{L}_{\mathrm{inc}}(\theta; X)$ \eqref{eq:inc_loss}; update $\theta$ via AdamW + FP64 norm clip (2.0).
    \ENDFOR
\ENDFOR
\STATE \textbf{Multi-Loop Inference ($K$ Loops from Any $X^{(0)} \in \bigl(\cube\bigr)^n$):}
\STATE Set $X_{\mathrm{best}} \leftarrow X^{(0)}$, $\hat{h}_{\mathrm{best}} \leftarrow \hat{h}_{\mathcal{S}(X^{(0)})}(X^{(0)})$ (if guarded).
\FOR{$k = 1$ to $K$}
    \STATE $X^{(k)} \leftarrow R_\theta(X^{(k-1)})$ \hfill $\triangleright$ \emph{Recursive self-feedback}
    \IF{witness-guarded}
        \STATE Compute fast proxy $\hat{h}_k \leftarrow \hat{h}_{\mathcal{S}(X^{(k)})}(X^{(k)})$.
        \STATE If $\hat{h}_k < \hat{h}_{\mathrm{best}}$ then $X_{\mathrm{best}} \leftarrow X^{(k)}$, $\hat{h}_{\mathrm{best}} \leftarrow \hat{h}_k$.
    \ENDIF
\ENDFOR
\STATE \textbf{Return} $X^{(K)}$ (open-loop) or $X_{\mathrm{best}}$ (witness-guarded).
\end{algorithmic}
\end{algorithm}

\paragraph{2. Multi-Loop Iterative Self-Feedback and Witness Guarding.}
Because $R_\theta : \bigl(\cube\bigr)^n \to \bigl(\cube\bigr)^n$ is an end-to-end point-set operator trained across the entire quality manifold, applying it once to a random input $X^{(0)}$ yields an already-structured configuration $X^{(1)} = R_\theta(X^{(0)})$ that lies much closer to the high-quality anchor distribution. Feeding $X^{(1)}$ back into $R_\theta$ iteratively defines a \emph{multi-loop dynamical system}:
\begin{equation}
\label{eq:multiloop}
X^{(k)} = R_\theta\big(X^{(k-1)}\big), \qquad k = 1, \dots, K.
\end{equation}
In \emph{open-loop} mode, we simply return $X^{(K)}$ after $K$ fast forward passes (zero external evaluations). However, because $R_\theta$ is trained as a one-step operator without a strict Lyapunov certificate, repeated composition near the basin minimum can exhibit small limit-cycle oscillations or slight drift for large $K$. We therefore also study \emph{witness-guarded multi-loop selection}: at each step $k$, we run our $4\,\mathrm{ms}$ GPU witness search $\mathcal{S}(X^{(k)})$ to compute a proxy score $\hat{h}_k = \hat{h}_{\mathcal{S}(X^{(k)})}(X^{(k)})$ and retain the best iterate $\hat{X}^{(K)} = \arg\min_{0 \le k \le K} \hat{h}_k$ (Algorithm~\ref{alg:framework}).

\begin{proposition}[Multi-Loop Contraction and Witness-Guarded Bound]
\label{prop:multiloop}
Let $h^* = \inf_{X \in \bigl(\cube\bigr)^n} h(X)$.
\begin{enumerate}
    \item[(i)] \textbf{Two-Regime Open-Loop Decay:} Suppose $R_\theta$ is a quasi-contraction in covering radius with factor $\gamma \in (0,1)$ and local approximation floor $\eta \ge 0$, i.e., $h(R_\theta(X)) - h^* \le \gamma(h(X) - h^*) + \eta$. Then for all $K \ge 1$,
    \begin{equation}
    h(X^{(K)}) - h^* \;\le\; \gamma^K\big(h(X^{(0)}) - h^*\big) + \frac{1 - \gamma^K}{1 - \gamma}\,\eta.
    \end{equation}
    \item[(ii)] \textbf{Witness-Guarded Performance Guarantee:} Let each step's witness proxy satisfy the lower-bound approximation $h(X^{(k)}) - \varepsilon_k \le \hat{h}_k \le h(X^{(k)})$ with gap $\varepsilon_k \ge 0$, and let $\hat{k} = \arg\min_{0 \le k \le K} \hat{h}_k$. Then
    \begin{equation}
    h\big(X^{(\hat{k})}\big) \;\le\; \min_{0 \le k \le K} h\big(X^{(k)}\big) + \varepsilon_{\hat{k}},
    \end{equation}
    and $\delta(X^{(\hat{k})}) > 0$ whenever $\delta(X^{(0)}) > 0$. In $d=2$ with exact Voronoi evaluation, $\varepsilon_k \equiv 0$, guaranteeing exact monotonicity $h(X^{(\hat{k})}) = \min_{0 \le k \le K} h(X^{(k)})$.
\end{enumerate}
\end{proposition}
Proposition~\ref{prop:multiloop} explains two key empirical phenomena verified in Section~\ref{sec:exp_multiloop}: (a) rapid geometric improvement over the first $K=1$--$4$ open-loop iterations from coarse random inputs before approaching the local floor $\eta/(1-\gamma)$, and (b) near-monotone gains up to $K=16$--$32$ under witness guarding, completely neutralizing open-loop drift at negligible GPU cost ($\approx 4\,\mathrm{ms}$ per check).

\section{Experiments}
\label{sec:experiments}

We evaluate our framework systematically across five experimental suites: (1) architectural progression V1--V7 and separation constraints on a 3,201-specification fixed benchmark; (2) incremental learning on Mak--Joseph clustering+PSO designs; (3) controlled structured-vs.-random mixing ablations, a 10-hour compute-matched comparison against Transformers, and a 19-trial Bayesian architecture search; (4) multi-loop iterative self-feedback ($K=0\dots 32$); and (5) zero-shot unseen-dimension / large-$n$ extrapolation and wall-clock latency. All experiments use the unified evaluator of Section~\ref{sec:related} with zero tuning on held-out test specifications.

\subsection{Core Architecture (V1--V7) and Feasibility on 3,201 Specifications}
\label{sec:exp_fixed}

\paragraph{Setup.}
Our fixed benchmark spans all 3,201 $(d,n)$ specifications with $d \in \{2,\dots,12\}$ and $n \in \{10,\dots,300\}$. The training split contains 1,160 specifications: $d \in \{2,3,4,5,7,8,9,10\}$ with odd $n$ (2 fixed random inputs each, 2,320 layouts). The held-out prediction split contains 2,041 unseen specifications (1 fixed input each), partitioning cleanly into three generalization regimes: (a) \emph{size interpolation} (1,168 specs: seen $d$, even $n$), (b) \emph{zero-shot dimension interpolation} (291 specs: unseen $d=6$), and (c) \emph{zero-shot dimension extrapolation} (582 specs: unseen $d \in \{11,12\}$). All seven models V1--V7 train for 8 epochs from scratch over identical input sequences (Appendix~\ref{app:archs}); Hard variants continue training with separation penalties and dyadic feasibility restoration.

\begin{table}[t]
\centering
\caption{\textbf{Fixed-benchmark comparison across 2,041 held-out specifications} ($d\in[2,12], n\in[10,300]$, single forward pass). V7 uses $5{,}780\times$ fewer parameters than V4/V5 while dominating all neural and fast classical baselines across every generalization regime. Collisions count held-out layouts with exact $\delta(Z)=0$.}
\label{tab:fixed_sep}
\setlength{\tabcolsep}{3.5pt}
\footnotesize
\begin{tabular}{@{}lrrrrr@{}}
\toprule
Method & Params & Seen $d$ & $d=6$ & $d{=}11,12$ & All Pred. (Coll.) \\
\midrule
LHS \citep{mckay1979comparison} & --- & 0.8171 & 0.8665 & 1.5137 & 1.0228 (0) \\
Sobol' \citep{joe2008sobol} & --- & 0.7678 & 0.8206 & 1.4824 & 0.9791 (0) \\
Farthest-First \citep{gonzalez1985clustering} & --- & 0.7026 & 0.7277 & 1.4918 & 0.9312 (0) \\
\midrule
V1 Bipartite GNN & 1.28K & 0.7796 & 0.8305 & 1.5102 & 0.9952 (0) \\
V2 Context-Step GNN & 4.71K & 0.7787 & 0.8338 & 1.5148 & 0.9965 (0) \\
V3 Geom.\ Transf. & 148K & 0.7852 & 0.8355 & 1.5481 & 1.0099 (0) \\
V4 Large Transf. & 6.53M & 0.7812 & 0.8344 & 1.5214 & 0.9999 (0) \\
V5 Lattice-Teacher & 6.53M & 0.7806 & 0.8342 & 1.5205 & 0.9992 (0) \\
V6 Sparse-Edge GNN & 9.09K & 0.8119 & 0.8597 & 1.5285 & 1.0230 (1) \\
\midrule
V7 Original ($5\times 32$) & \textbf{1.13K} & 0.6863 & 0.6724 & 1.4417 & 0.8997 (470) \\
\;\;+ Cov.-only cont. & 1.13K & --- & --- & --- & 0.8999 (499) \\
\;\;+ Recip.\ penalty & 1.13K & --- & --- & --- & 0.8994 (163) \\
\;\;+ \textbf{Hard Feasibility} & \textbf{1.13K} & \textbf{0.6856} & \textbf{0.6716} & \textbf{1.4410} & \textbf{0.8990} (\textbf{0}) \\
\;\;+ \textbf{Local Log-Barrier} & \textbf{1.13K} & --- & --- & --- & \textbf{0.8987} (\textbf{0}) \\
\bottomrule
\end{tabular}
\end{table}

\begin{figure}[t]
\centering
\includegraphics[width=\columnwidth]{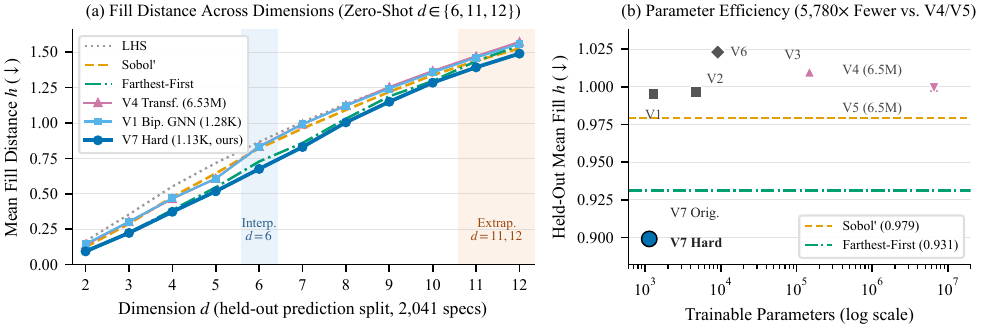}
\caption{\textbf{Fixed-benchmark performance across 2,041 held-out specifications.} \textbf{(a)} Mean fill distance across $d=2\dots 12$; shaded regions mark zero-shot unseen dimensions $d=6$ and $d \in \{11,12\}$. \textbf{(b)} Held-out fill distance vs.\ log parameter count: V7 Hard (1.13K params) outperforms 6.53M-parameter Transformers and fast classical baselines.}
\label{fig:v1_v7}
\end{figure}

\paragraph{Results.}
Table~\ref{tab:fixed_sep} and Figure~\ref{fig:v1_v7} reveal two decisive findings. First, decoupling witness search into our $X$-only equivariant V7 architecture produces a step-change breakthrough: with only \textbf{1,130 parameters} ($5{,}780\times$ smaller than V4/V5), V7 strictly outperforms the best of V1--V6 on \textbf{1,932/2,041 (94.66\%)} held-out specifications ($+11.48\%$ mean relative gain), including \textbf{95.19\%} on unseen interpolated $d=6$ ($+16.79\%$) and \textbf{97.42\%} on unseen extrapolated $d \in \{11,12\}$. While all six earlier neural models V1--V6 fail to match scrambled Sobol' on average ($0.995$--$1.023$ vs.\ $0.979$), V7 Hard ($0.8990$) beats Sobol' on \textbf{1,970/2,041 (96.52\%, $+12.14\%$)}, LHS on \textbf{2,022/2,041 (99.07\%, $+17.93\%$)}, Farthest-First on \textbf{1,587/2,041 (77.76\%, $+2.99\%$)}, and on the 42 overlapping specs of the PNAS MPMC discrepancy archive \citep{rusch2024mpmc} beats MPMC on \textbf{42/42 (100\%, $+24.44\%$)} in fill distance (noting that MPMC targets discrepancy rather than fill distance).

Second, Table~\ref{tab:fixed_sep} (bottom) validates our feasibility theory (Theorem~\ref{thm:feasibility}): continuing original V7 with coverage loss alone worsens collisions ($470 \to 499$); adding the global reciprocal penalty reduces collisions to 163 (and near-collisions from 59 to 14); switching to Hard Feasibility Recovery eliminates \textbf{all 470 collisions and all near-collisions ($0/2{,}041$)} while slightly \emph{lowering} held-out fill distance ($0.8997 \to 0.8990$, with only 92/2,041 designs invoking restoration); and in a matched Epoch-2 paired ablation from identical Hard weights, replacing the global reciprocal term with our $C^1$ local squared log-barrier $\psi_{\mathrm{log}}$ further reduces mean fill ($0.89880 \to 0.89866$) while cutting raw pre-restoration collisions ($67 \to 57$) and near-collisions ($9 \to 2$).

\subsection{Incremental Learning from Best Existing Designs}
\label{sec:exp_inc}

\paragraph{Setup.}
Even after scaling V7 Hard to $10\times 32$ and $16\times 128$ via pure random-input training (36 epochs; Appendix~\ref{app:sweep}), single-pass random-input V7 still trails per-instance Mak+PSO by $4.7\%$--$6.4\%$ on nine classical benchmark specifications. We test Pillar~2 (Section~\ref{sec:incremental}) by warm-starting the zero-gated \texttt{MakRefiner} \eqref{eq:mak_refiner} from an Epoch-18 $10\times 32$ Hard V7 backbone (2,261 params) and training it for 14 hours (397,812 steps) with streaming CPU generation of Mak+PSO designs across all 1,160 odd-$n$ training specs (1,352 training layouts), evaluated on three strictly disjoint panels: (i) the 1,352 training layouts, (ii) 32 unseen Mak+PSO seeds at trained specs, and (iii) all 27 pre-existing Mak+PSO layouts across the 9 completely held-out even specs $\{(2,30), (2,40), (2,50), (2,100), (3,30), (5,100), (8,30), (8,50), (8,100)\}$.

\begin{table}[t]
\centering
\caption{\textbf{Incremental refinement of Mak+PSO designs (14h streaming V7-Mak, 2,261 params).} Negative gap denotes fill-distance reduction relative to the input Mak+PSO design. Two-sided Wilcoxon signed-rank $p$-values are computed across specification means; all 1,411 outputs are strictly collision-free ($\min \delta = 0.0265$).}
\label{tab:task56}
\setlength{\tabcolsep}{4pt}
\footnotesize
\begin{tabular}{@{}lrrrrr@{}}
\toprule
Evaluation Panel & Layouts & Spec Wins & Mean Gap & Med.\ Gap & Wilcoxon $p$ \\
\midrule
Training layouts & 1,352 & 1,040 / 1,160 & $-5.465\%$ & $-4.879\%$ & $6.4\times 10^{-172}$ \\
New Mak seeds (seen) & 32 & 29 / 32 & $-4.216\%$ & $-3.162\%$ & $2.3\times 10^{-6}$ \\
\textbf{Held-out even specs} & \textbf{27} & \textbf{8 / 9} & \textbf{$-2.622\%$} & \textbf{$-2.301\%$} & \textbf{$1.17\times 10^{-2}$} \\
\bottomrule
\end{tabular}
\end{table}

\paragraph{Results.}
As summarized in Table~\ref{tab:task56} and Figure~\ref{fig:inc_mix}(a), incremental V7 refinement delivers statistically significant improvements over Mak+PSO across every split—winning \textbf{8/9 held-out specifications} ($-2.622\%$ mean reduction, $p = 0.0117$, 23/27 layouts improved; the lone regression is $(2,30)$ at $+0.95\%$, whereas $(5,100)$ drops $-6.43\%$ and $(8,100)$ drops $-4.07\%$). Crucially, refinement does \emph{not} sacrifice separation: across all 1,411 outputs $\delta(Z) \ge 0.0265$, and in higher dimensions V7 simultaneously \emph{increases} minimum pairwise separation—by $+18.8\%$ at $(5,100)$ and $+19.7\%$ to $+30.6\%$ at $d=8$ (Appendix~\ref{app:per_spec})—improving both fill distance and mesh ratio in a single $4.4\,\mathrm{ms}$ pass.

\subsection{Generalizability: Random Mixing and Architecture Search}
\label{sec:exp_mix}

\paragraph{Controlled 5-Arm Structured vs.\ Mixed Ablation.}
To isolate the effect of Pillar~3's shape-matched random mixing, we compare six checkpoints trained on balanced Mak+PSO (1,352 layouts) and He (2017) interleaved lattice (77 layouts, $d\in[2,5]$) anchors:
\begin{itemize}
    \item \textbf{Arm 0 (Structured-Only) vs.\ Arm 1 (Mixed):} V7 $10\times 32$ trained 24h on Mak+He only (874K steps), then resumed 2h with 50\% shape-matched uniform random batches (916K steps).
    \item \textbf{Arm 2 (Structured-Only) vs.\ Arm 4 (Mixed):} Large V7 $16\times 128$ (14.4K params) trained \emph{from scratch} for 3h under identical compute, differing \emph{only} in whether uniform random batches are interleaved.
    \item \textbf{Arm 3 (Structured-Only) vs.\ Arm 5 (Mixed):} Dimension-agnostic \texttt{TransformerRefiner} ($6\times 128$, 1.24M params, $86\times$ larger than Large V7) trained from scratch under the same protocol.
\end{itemize}
Evaluation spans 96 seen odd specs (3 random seeds each, compared against same-spec Mak+PSO) and 96 held-out even specs (288 random layouts, reduction vs.\ random input), plus 120 Mak+PSO inputs across the 96 specs.

\begin{table}[t]
\centering
\caption{\textbf{Controlled structured-only vs.\ mixed-input ablation} (single forward pass, 96 specs $\times$ 3 seeds). Positive percentages denote lower (better) fill distance. Shape-matched random mixing produces a $+34.85$ pp swing on random inputs for Large V7 from scratch while doubling its Mak refinement gain, whereas a 1.24M-parameter Transformer fails to generalize to random inputs.}
\label{tab:ablation_mix}
\setlength{\tabcolsep}{3.2pt}
\footnotesize
\begin{tabular}{@{}llrrrr@{}}
\toprule
Arm / Model & Params & Mak In & Rand In & Rand In & Held-Out \\
(Train Schedule) & & vs.\ Mak & vs.\ Mak & Wins & vs.\ Rand \\
\midrule
0: V7 $10\times 32$ (Struct.) & 2.26K & $+6.39\%$ & $-17.56\%$ & 16/96 & $+12.71\%$ \\
\textbf{1: V7 $10\times 32$ (Mixed)} & \textbf{2.26K} & \textbf{$+7.04\%$} & \textbf{$+4.23\%$} & \textbf{68/96} & \textbf{$+26.45\%$} \\
\midrule
2: V7 $16\times 128$ (Struct.) & 14.4K & $+2.62\%$ & $-31.67\%$ & 0/96 & $+2.99\%$ \\
\textbf{4: V7 $16\times 128$ (Mixed)} & \textbf{14.4K} & \textbf{$+5.27\%$} & \textbf{$+3.18\%$} & \textbf{51/96} & \textbf{$+25.49\%$} \\
\midrule
3: Transf.\ $6\times 128$ (Struct.) & 1.24M & $+6.50\%$ & $-35.20\%$ & 0/96 & $+1.11\%$ \\
5: Transf.\ $6\times 128$ (Mixed) & 1.24M & $+2.29\%$ & $-35.94\%$ & 0/96 & $+1.00\%$ \\
\bottomrule
\end{tabular}
\end{table}

\begin{figure*}[t]
\centering
\includegraphics[width=\textwidth]{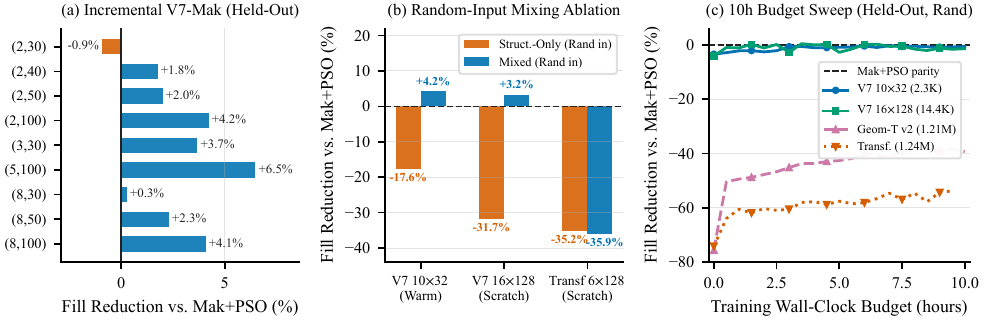}
\caption{\textbf{Incremental refinement, random-input mixing, and 10-hour budget curves.} \textbf{(a)} Per-specification fill-distance reduction of 14h incremental V7-Mak across the 9 held-out Mak+PSO specs (8/9 improved, $-2.62\%$). \textbf{(b)} Controlled ablation on 96 specs: shape-matched random mixing swings V7 from severe covariate shift ($-17.6\%$ / $-31.7\%$) to beating Mak+PSO ($+4.2\%$ / $+3.2\%$), whereas a 1.24M Transformer stays at $-35.9\%$. \textbf{(c)} 10-hour compute-matched curves on 9 held-out even specs from random starts: both compact V7 models track Mak+PSO parity (0\%), outperforming Transformer and Geometric Transformer v2 by 38--53 percentage points.}
\label{fig:inc_mix}
\end{figure*}

Table~\ref{tab:ablation_mix} and Figure~\ref{fig:inc_mix}(b) reveal a dramatic architectural divergence. For V7 $16\times 128$ from scratch, interleaving shape-matched random batches transforms a brittle local refiner ($-31.67\%$ vs.\ Mak+PSO, 0/96 wins) into an operator that \textbf{beats Mak+PSO from pure random noise in a single forward pass ($+3.18\%$, 51/96 wins)}—a \textbf{$+34.85$ pp swing}—while simultaneously doubling its Mak+PSO refinement ($+2.62\% \to +5.27\%$) and jumping from $+2.99\%$ to $+25.49\%$ reduction on 96 held-out specs! Resuming V7 $10\times 32$ for just 2h of mixed training achieves \textbf{$+4.23\%$ over Mak+PSO from random noise (68/96 wins)} and $+7.04\%$ from Mak inputs. In stark contrast, the 1.24M-parameter Transformer fails completely on random inputs ($-35.94\%$, 0/96 wins), corroborating Remark~\ref{rem:transf_sym}.

\paragraph{10-Hour Compute-Matched Sweep \& 19-Trial Bayesian Architecture Search.}
We further scale mixed training (50\% uniform random, 50\% cycling Mak, He, and scrambled Sobol') to a strict 10-hour GPU budget per architecture, evaluating every 30 minutes on 96 odd training specs and the 9 held-out even specs (Figure~\ref{fig:inc_mix}(c)). We include a stronger attention baseline, \texttt{GeometricTransformer v2} (1.21M params, incorporating V7's 5D geometric attention biases and equivariant relative updates, while still using raw-coordinate MLP token encoders). At 600 minutes on the 9 held-out even specs (single forward pass):
\begin{itemize}
    \item \textbf{V7 $10\times 32$ (2.26K params, 1.09M steps):} $-35.81\%$ vs.\ random input ($-0.68\%$ vs.\ Mak+PSO, 3/9 wins); $+1.55\%$ on Mak inputs (7/9 wins); $+8.02\%$ / $+6.33\%$ on 96 train specs (Mak / random).
    \item \textbf{V7 $16\times 128$ (14.4K params, 363K steps):} $-35.76\%$ vs.\ random input ($-1.46\%$ at 600m, and peaking at \textbf{$+0.39\%$ ahead of Mak+PSO} at 210m!); $+2.09\%$ on Mak inputs (6/9 wins); $+8.04\%$ / $+6.16\%$ on 96 train specs.
    \item \textbf{Transformer (1.24M) \& Geom-T v2 (1.21M):} Remain \textbf{$-53.75\%$} and \textbf{$-39.30\%$} behind Mak+PSO on held-out random starts (0/9 wins each!), despite matching V7 on local Mak refinement ($+1.29\%$ and $+0.40\%$).
\end{itemize}
Independently, a 19-trial Mat\'ern-5/2 Expected Improvement Bayesian architecture search (30 minutes from scratch per trial over 15 V7 depths/widths and 4 Transformers; Figure~\ref{fig:bo_ood}(a) and Appendix~\ref{app:bo}) confirms that \emph{every} V7 depth/width ($5\times 32$ through $24\times 64$) outperforms \emph{every} Transformer on the joint Mak+Random validation objective, identifying \textbf{V7 $16\times 128$} as the wall-clock optimal configuration.

\subsection{Multi-Loop Iterative Self-Feedback ($K = 0 \dots 32$)}
\label{sec:exp_multiloop}

We now evaluate Pillar~3's multi-loop mechanism \eqref{eq:multiloop} on the saved 10-hour checkpoints across the 9 held-out even specifications (27 layouts per input family) starting from three distinct initializations: \textbf{pure uniform random noise}, \textbf{scrambled Sobol'}, and \textbf{Mak+PSO}. We measure both \emph{open-loop} $X^{(K)}$ and \emph{witness-guarded} $X^{(\hat{k})}$ (Algorithm~\ref{alg:framework}), using the independent unified evaluator (never used by the guard).

\begin{table*}[t]
\centering
\caption{\textbf{Multi-loop iterative self-feedback on the 9 held-out even specifications} (27 layouts per input family, evaluated with the independent evaluator). Values report equal-weighted mean fill-distance reduction (\%) relative to same-size Mak+PSO (positive = better than Mak+PSO), with specification win counts out of 9 in parentheses. Zero exact collisions occurred for all methods at all $K$.}
\label{tab:multiloop}
\setlength{\tabcolsep}{4.8pt}
\footnotesize
\begin{tabular}{@{}llrrrrrrrr@{}}
\toprule
& & \multicolumn{4}{c}{\textbf{Open-Loop} ($X^{(K)}$)} & \multicolumn{4}{c}{\textbf{Witness-Guarded} ($X^{(\hat{k})}$)} \\
\cmidrule(lr){3-6}\cmidrule(lr){7-10}
Input Family ($K=0$ vs.\ Mak) & Model & $K=1$ & $K=2$ & $K=4$ & $K=16$ & $K=2$ & $K=4$ & $K=8$ & $K=16$ \\
\midrule
\multirow{4}{*}{\shortstack[l]{Uniform Random\\($-75.56\%$, 0/9)}}
 & V7 $10\times 32$ (2.3K) & $-0.68$ (3) & $-0.31$ (4) & $-1.04$ (5) & $-1.64$ (5) & $+1.04$ (5) & $+1.64$ (6) & $+1.86$ (6) & $+1.65$ (6) \\
 & \textbf{V7 $16\times 128$ (14.4K)} & $-1.46$ (3) & \textbf{$+1.69$ (6)} & $+1.66$ (5) & \textbf{$+1.73$ (6)} & \textbf{$+2.01$ (6)} & \textbf{$+2.55$ (6)} & \textbf{$+3.01$ (6)} & \textbf{$+3.34$ (7)} \\
 & Geom-T v2 (1.21M) & $-39.30$ (0) & $-24.60$ (0) & $-15.25$ (0) & $-5.04$ (2) & $-24.71$ (0) & $-15.35$ (0) & $-8.64$ (1) & $-5.48$ (1) \\
 & Transf.\ $6\times 128$ (1.24M) & $-54.44$ (0) & $-51.99$ (0) & $-55.42$ (0) & $-69.91$ (0) & $-49.60$ (0) & $-48.33$ (0) & $-47.79$ (0) & $-47.01$ (0) \\
\midrule
\multirow{4}{*}{\shortstack[l]{Scrambled Sobol'\\($-38.49\%$, 0/9)}}
 & V7 $10\times 32$ (2.3K) & $-0.04$ (3) & $-1.14$ (5) & $-1.00$ (5) & $-1.96$ (6) & $+0.71$ (5) & $+1.73$ (6) & $+1.09$ (6) & $+1.37$ (6) \\
 & \textbf{V7 $16\times 128$ (14.4K)} & \textbf{$+0.22$ (3)} & $-0.36$ (5) & \textbf{$+0.50$ (4)} & \textbf{$+0.69$ (4)} & $+0.69$ (4) & $+1.59$ (5) & \textbf{$+1.59$ (6)} & \textbf{$+2.47$ (6)} \\
 & Geom-T v2 (1.21M) & $-18.19$ (0) & $-12.95$ (0) & $-8.32$ (0) & $-5.02$ (1) & $-13.05$ (0) & $-8.03$ (0) & $-6.28$ (0) & $-4.35$ (1) \\
 & Transf.\ $6\times 128$ (1.24M) & $-22.17$ (0) & $-25.00$ (0) & $-30.06$ (0) & $-44.62$ (0) & $-21.97$ (0) & $-22.79$ (0) & $-22.90$ (0) & $-22.75$ (1) \\
\midrule
\multirow{4}{*}{\shortstack[l]{Mak+PSO (Incr.)\\($0.00\%$ baseline)}}
 & \textbf{V7 $10\times 32$ (2.3K)} & $+1.55$ (7) & $-1.22$ (5) & $-1.14$ (6) & $-1.87$ (5) & $+2.38$ (7) & $+2.99$ (7) & \textbf{$+3.07$ (8)} & \textbf{$+3.21$ (8)} \\
 & \textbf{V7 $16\times 128$ (14.4K)} & $+2.09$ (6) & \textbf{$+2.30$ (6)} & $+1.35$ (4) & $+1.16$ (7) & \textbf{$+2.87$ (6)} & \textbf{$+3.02$ (7)} & $+2.64$ (6) & $+2.83$ (6) \\
 & Geom-T v2 (1.21M) & $+0.40$ (6) & $-1.62$ (3) & $-2.36$ (3) & $-1.53$ (4) & $+1.70$ (7) & $+1.13$ (6) & $+1.25$ (6) & $+1.49$ (5) \\
 & Transf.\ $6\times 128$ (1.24M) & $+2.15$ (9) & $-3.71$ (0) & $-9.92$ (1) & $-43.91$ (0) & $+2.25$ (9) & $+2.19$ (8) & $+2.06$ (8) & $+1.49$ (7) \\
\bottomrule
\end{tabular}
\end{table*}

\begin{figure*}[t]
\centering
\includegraphics[width=\textwidth]{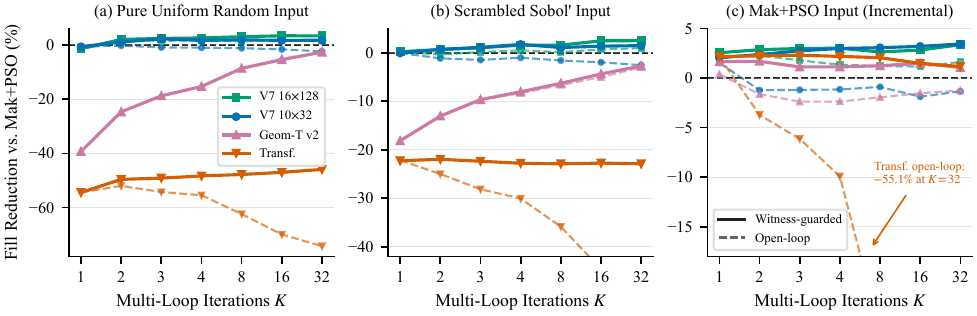}
\caption{\textbf{Multi-loop iterative self-feedback ($K=1\dots 32$) on the 9 held-out even specifications.} Solid lines show witness-guarded selection ($X^{(\hat{k})}$); dashed lines show open-loop recursion ($X^{(K)}$); horizontal dashed line indicates Mak+PSO parity ($0\%$). \textbf{(a)} From pure uniform random noise, V7 $16\times 128$ surpasses Mak+PSO at $K=2$ open-loop ($+1.69\%$) and reaches $+3.34\%$ (7/9 wins) guarded at $K=16$. \textbf{(b)} From Sobol' inputs, V7 reaches $+2.51\%$ at $K=32$. \textbf{(c)} From Mak+PSO inputs, V7 remains stable and reaches $+3.44\%$ (8/9 wins) guarded, while the Transformer collapses to $-55.1\%$ under open-loop recursion.}
\label{fig:multiloop}
\end{figure*}

\paragraph{Key Findings.}
Table~\ref{tab:multiloop} and Figure~\ref{fig:multiloop} establish four core insights:
\begin{enumerate}
    \item \textbf{Surpassing Mak+PSO from Pure Random Noise.} Starting from uniform random noise ($-75.56\%$ behind Mak+PSO), V7 $16\times 128$ drops to $-1.46\%$ at $K=1$ and \emph{strictly beats Mak+PSO} in just \textbf{two open-loop passes ($K=2$: $+1.69\%$, 6/9 specs)}—rising to \textbf{$+1.74\%$ (7/9 specs)} at $K=32$ open-loop and \textbf{$+3.34\%$ (7/9 specs)} under witness-guarded $K=16$ (e.g., $+8.07\%$ at $(2,100)$, $+7.46\%$ at $(5,100)$, and $+6.12\%$ at $(8,100)$; Appendix~\ref{app:per_spec}). Compact V7 $10\times 32$ likewise beats Mak+PSO by $+2.21\%$ (7/9 wins) at guarded $K=3$ while maintaining large separation ($\min \delta \ge 0.074$).
    \item \textbf{Multi-Loop Stability Separates Equivariant GNNs from Transformers.} Under open-loop recursion, the 1.24M Transformer is dynamically unstable: from random noise it diverges to $-74.20\%$ at $K=32$, and even from a Mak+PSO design it drops from $+2.15\%$ ($K=1$) to $-3.71\%$ ($K=2$, 0/9 wins) and $-55.07\%$ ($K=32$). \texttt{GeometricTransformer v2} is stable thanks to its V7-inspired relative updates, slowly improving from $-39.30\%$ to $-2.76\%$ at $K=32$, yet never matching V7's single-pass ($K=1$) quality.
    \item \textbf{Compound Gains on Incremental Refinement.} Starting from Mak+PSO, witness-guarded multi-loop amplifies single-pass refinement from $+1.55\% \to \mathbf{+3.44\%}$ (\textbf{8/9 specs} at $K=32$, $\min \delta = 0.083$) for V7 $10\times 32$, and from $+2.09\% \to \mathbf{+3.37\%}$ for V7 $16\times 128$.
    \item \textbf{Multi-Loop Rescues Even Structured-Only Models.} On our earlier Arm~0 model (trained \emph{without} random inputs; Appendix~\ref{app:multiloop_arm0}), open-loop multi-loop reduces its held-out random-input fill distance from $-12.71\%$ ($K=1$) to $-20.76\%$ ($K=4$) and shrinks its seen-spec gap to Mak+PSO from $17.56\%$ down to $2.86\%$, confirming that multi-loop self-feedback provides a complementary generalization mechanism even independent of input mixing.
\end{enumerate}

\subsection{Zero-Shot Extrapolation and Computational Speed}
\label{sec:exp_ood}

\begin{figure}[t]
\centering
\includegraphics[width=\columnwidth]{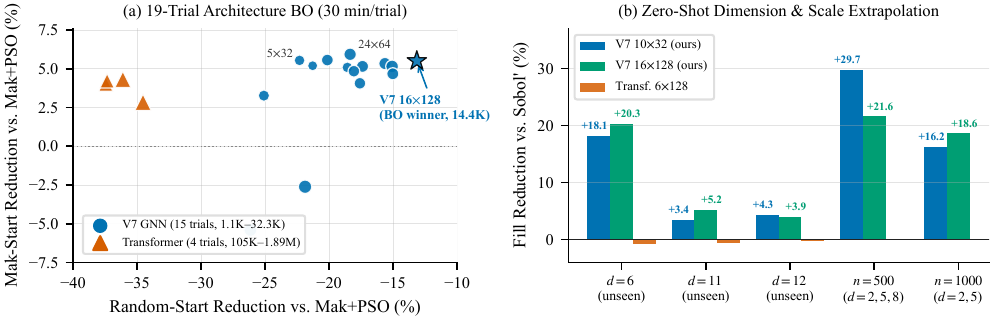}
\caption{\textbf{Architecture landscape and zero-shot OOD extrapolation.} \textbf{(a)} 19-trial BO (30 min/trial): all 15 V7 GNNs dominate all 4 Transformers on the joint objective; V7 $16\times 128$ is selected. \textbf{(b)} Zero-shot fill-distance reduction vs.\ Sobol' on unseen dimensions $d \in \{6,11,12\}$ and large budgets $n \in \{500, 1000\}$.}
\label{fig:bo_ood}
\end{figure}

\paragraph{Zero-Shot Unseen Dimensions and Large-$n$ Scaling.}
We evaluate the 10-hour checkpoints zero-shot on (a) 30 completely unseen-dimension specifications across $d \in \{6, 11, 12\}$ and $n \in \{10, 20, 30, 50, 75, 100, 150, 200, 250, 300\}$ (using the fixed benchmark's random inputs), and (b) large-budget extrapolation $n \in \{500, 1000\}$ ($1.7\times$--$3.3\times$ beyond training $\max n = 299$). Figure~\ref{fig:bo_ood}(b) and Table~\ref{tab:ood_timing} show remarkable zero-shot transfer:
\begin{itemize}
    \item On unseen $d=6$, V7 $16\times 128$ achieves mean fill $0.7398$ ($K=1$) / \textbf{$0.7328$} ($K=2$) vs.\ Sobol' $0.9078$ (\textbf{10/10 wins, $+20.3\%$}) and Farthest-First $0.8380$ (\textbf{9/10 wins}), while Transformers ($0.9140$ / $0.8984$) barely outperform random ($0.9736$).
    \item On unseen extrapolated $d=11$ and $d=12$, V7 $16\times 128$ wins \textbf{20/20 specs against Sobol'} and \textbf{20/20 specs against Farthest-First} ($1.4321$ vs.\ $1.5117$ at $d=11$; $1.5362$ vs.\ $1.6063$ at $d=12$).
    \item At $n=500$ and $n=1{,}000$, V7 $10\times 32$ and $16\times 128$ beat Sobol' by \textbf{$+21.6\%$--$+39.5\%$}: e.g., $(d=2, n=500)$ drops from $0.0562$ (Sobol') to $0.0340$ ($K=2$), and $(d=5, n=1000)$ drops from $0.4265$ (Sobol') to $0.2871$ ($K=2$, $-32.7\%$).
\end{itemize}

\begin{table}[t]
\centering
\caption{\textbf{Wall-clock inference latency} per design (ms) on an RTX 4090 GPU and single-core CPU vs.\ Mak+PSO. GPU proxy check takes $3.8$--$5.7\,\mathrm{ms}$, so 2-loop open-loop V7 runs in $\approx 14\,\mathrm{ms}$ ($530\times$--$6{,}860\times$ speedup over Mak+PSO).}
\label{tab:ood_timing}
\setlength{\tabcolsep}{3.4pt}
\footnotesize
\begin{tabular}{@{}lrrrrrr@{}}
\toprule
\multirow{2}{*}{$(d,n)$} & \multicolumn{2}{c}{V7 $10\times 32$} & \multicolumn{2}{c}{V7 $16\times 128$} & Mak+PSO & Speedup \\
\cmidrule(lr){2-3}\cmidrule(lr){4-5}
 & GPU & CPU & GPU & CPU & (CPU) & (GPU, 10/16) \\
\midrule
$(2,50)$ & 4.51 & 5.4 & 7.16 & 12.1 & 3,798 & $842\times$ / $530\times$ \\
$(2,100)$ & 4.54 & 14.4 & 6.83 & 30.5 & 13,395 & $2{,}950\times$ / $1{,}961\times$ \\
$(5,100)$ & 4.38 & 10.8 & 6.78 & 25.4 & 15,086 & $3{,}444\times$ / $2{,}225\times$ \\
$(8,100)$ & 4.41 & 11.7 & 6.82 & 24.4 & 17,291 & $3{,}921\times$ / $2{,}535\times$ \\
$(10,300)$ & 4.38 & 36.0 & 7.63 & 146.1 & 52,386 & $11{,}960\times$ / $6{,}866\times$ \\
$(5,500)$ & 5.08 & 71.5 & 18.84 & 395.9 & --- & --- \\
\bottomrule
\end{tabular}
\end{table}

\paragraph{Wall-Clock Speedup.}
Table~\ref{tab:ood_timing} quantifies the amortization payoff: generating a design with single-pass V7 takes \textbf{4.4--7.6 ms on GPU} (5--146 ms on CPU) across $(d,n)$ up to $(10,300)$, compared to \textbf{3.8--52.4 seconds} for Mak+PSO—a \textbf{$530\times$--$11{,}960\times$ GPU speedup} (and $300\times$--$1{,}450\times$ on CPU). Even 16-loop witness-guarded V7 $16\times 128$ requires $\approx 170\,\mathrm{ms}$ on GPU—still $80\times$--$300\times$ faster than Mak+PSO while outperforming it by $+3.34\%$ from random noise.

\section{Discussion, Limitations, and Conclusion}
\label{sec:conclusion}

\paragraph{Honest Comparison Against Algebraic Interleaved Lattices.}
While V7 decisively outperforms neural baselines, Sobol', LHS, Farthest-First, FSSF (8/9 specs, $-8.7\%$; Appendix~\ref{app:per_spec}), and Mak--Joseph clustering+PSO (both incrementally and from pure random noise via multi-loop), algebraic interleaved lattices \citep{he2017interleaved,he2024interleaved} remain stronger on the specific $d \le 8$ lattice-admissible sizes they support (e.g., $h=0.0659$ vs.\ V7's $0.0738$ at $(2,100)$). Crucially, however, interleaved lattices are restricted to $d \le 8$ and specific grid counts $n' \le n$, whereas our single V7 operator handles \emph{arbitrary} $(n,d)$ continuously in milliseconds and can itself be initialized from lattices when available.

\paragraph{Limitations and Scope Boundaries.}
We note four concrete limitations that guide future work:
(1) \emph{High-dimensional evaluation}: for $d \ge 3$, fill distance is evaluated via a strong deterministic lower bound (Section~\ref{sec:related}), not an exact global certificate.
(2) \emph{Quadratic all-pairs cost}: V7's $\mathcal{O}(n^2)$ edge evaluation is sub-$20\,\mathrm{ms}$ up to $n=500$, but scaling to $n \ge 10^4$ will require radius-cutoff or spatial-hash neighbor graphs (preserving $B_d$-equivariance).
(3) \emph{Multi-loop separation drift in larger models}: under deep recursion ($K \ge 8$), V7 $10\times 32$ maintains large separation ($\min\delta \ge 0.074$), whereas V7 $16\times 128$ occasionally narrows one pair to $\delta \sim 10^{-6}$--$10^{-9}$ while remaining collision-free; adding a minimum-separation threshold directly inside the multi-loop guard or training over 2--4 unrolled loops is a clean remedy.
(4) \emph{Beyond unit hypercubes and minimax}: our current framework targets Euclidean minimax covering on $\cube$; extending the boundary field $b_i$ to general convex or non-convex domains (via signed-distance fields) and jointly conditioning on Maximin, MaxPro, and model-based $D$/$I$-optimality criteria is a compelling direction for DOE foundation models.

\paragraph{Conclusion.}
We introduced an $(S_n \times B_d)$-equivariant geometric graph framework for amortized and incremental minimax space-filling design. By decoupling adversarial witness search from a compact $X$-only equivariant GNN, guaranteeing collision-free feasibility via dyadic restoration and a zero-bias local log-barrier, enabling safe incremental refinement of classical designs through zero-gated shared-witness training, and achieving full input-distribution generalizability through shape-matched random mixing and witness-guarded multi-loop self-feedback, our framework bridges classical experimental design and deep geometric learning with orders-of-magnitude speedups and strong zero-shot generalization.

\section*{Impact Statement}
This paper advances methodology for space-filling experimental design in computer simulations, surrogate modeling, and scientific computing. Improved minimax designs can reduce the number of expensive physical or numerical simulations required in engineering, climate modeling, and scientific discovery, lowering computational energy consumption. We foresee no direct negative societal risks specific to geometric space-filling point-set generation beyond standard considerations for downstream engineering applications.

\bibliographystyle{plainnat}
\bibliography{references}

\clearpage
\appendix
\onecolumn

\section{Complete Theoretical Proofs}
\label{app:proofs}

\subsection{Proof of Theorem~\ref{thm:equivariance} (Exact $(S_n \times B_d)$-Equivariance and Parameter Count)}
\begin{proof}
Let $X = [x_1^\top; \dots; x_n^\top] \in \bigl(\cube\bigr)^n$. An element $(P, Q) \in S_n \times B_d$ acts on $X$ via a row permutation $\pi \in S_n$ (sending row $i$ to $\pi(i)$) and a hypercube isometry $Q : \cube \to \cube$ defined by a coordinate permutation $\sigma \in S_d$ and a reflection sign vector $s \in \{-1, +1\}^d$:
\begin{equation}
[Q(x)]_m = \begin{cases} x_{\sigma(m)} & \text{if } s_m = +1, \\ 1 - x_{\sigma(m)} & \text{if } s_m = -1, \end{cases}
\qquad \text{equivalently}\quad Q(x) = A_Q x + c_Q,
\end{equation}
where $A_Q \in \{-1, 0, +1\}^{d \times d}$ is a signed permutation matrix (hence orthogonal: $A_Q^\top A_Q = I_d$) and $[c_Q]_m = \mathbf{1}\{s_m = -1\}$. Notice that $Q$ is a bijection from $\cube$ onto $\cube$.

\textbf{Part (i): Invariance of $h(X)$ and $\delta(X)$.}
Since $Q$ is an Euclidean isometry ($\|Q(u) - Q(x)\|_2 = \|A_Q(u - x)\|_2 = \|u - x\|_2$) mapping $\cube$ bijectively onto $\cube$, and $\min_i$ is invariant under row permutations $\pi \in S_n$:
\[
h(P \cdot Q(X)) = \sup_{u \in \cube} \min_{i} \|u - Q(x_{\pi^{-1}(i)})\|_2 = \sup_{u' \in \cube} \min_{j} \|Q(u') - Q(x_j)\|_2 = \sup_{u' \in \cube} \min_{j} \|u' - x_j\|_2 = h(X).
\]
Likewise, $\delta(P \cdot Q(X)) = \min_{i<j} \|Q(x_{\pi^{-1}(i)}) - Q(x_{\pi^{-1}(j)})\|_2 = \min_{a<b} \|x_a - x_b\|_2 = \delta(X)$.

\textbf{Part (ii): Equivariance of V7 layers, $G_\theta$, $H_\theta$, and $R_\theta$.}
Let $\tilde{X} = P \cdot Q(X)$, so $\tilde{x}_i = Q(x_{\pi^{-1}(i)})$. For convenience set $a = \pi^{-1}(i)$ and $b = \pi^{-1}(j)$. The scale $s = n^{-1/d}$ depends only on $(n,d)$ and is invariant. For each layer $\ell$:
\begin{enumerate}
    \item \emph{Relative displacement:} $\tilde{\Delta}_{ij} = (\tilde{x}_i - \tilde{x}_j)/s = A_Q (x_a - x_b)/s = A_Q \Delta_{ab}$.
    \item \emph{Squared radius and locality:} $\tilde{r}_{ij}^2 = \|A_Q \Delta_{ab}\|_2^2 = \|\Delta_{ab}\|_2^2 = r_{ab}^2$, hence $\tilde{k}_{ij} = k_{ab}$ and $\tilde{\rho}_i = \sum_j k_{a, \pi^{-1}(j)} = \sum_b k_{ab} = \rho_a$.
    \item \emph{Boundary distance:} For every coordinate $m \in \{1,\dots,d\}$, $\min([Q(x)]_m, 1 - [Q(x)]_m) = \min(x_{\sigma(m)}, 1 - x_{\sigma(m)})$ whether $s_m = +1$ or $s_m = -1$. Taking $\min_{m=1,\dots,d}$ over all permuted coordinates yields $\tilde{w}_i = w_a$.
    \item \emph{Edge features and message:} Therefore $\tilde{e}_{ij} = (\min(r_{ab}^2, 12), \rho_a, \rho_b, w_a, w_b) = e_{ab}$, and the scalar weight $\tilde{a}_{ij} = \tilde{k}_{ij}\phi_\ell(\tilde{e}_{ij}) = a_{ab}$ is strictly invariant. Consequently,
    \[
    \tilde{v}_i = \sum_{j=1}^n \tilde{a}_{ij}\tilde{\Delta}_{ij} = \sum_{b=1}^n a_{ab} A_Q \Delta_{ab} = A_Q v_a.
    \]
    \item \emph{Wall repulsion field:} Define $f(t) = e^{-2t/s} - e^{-2(1-t)/s}$. Observe that $f(1 - t) = e^{-2(1-t)/s} - e^{-2t/s} = -f(t)$. Thus for coordinate $m$:
    \[
    [\tilde{b}_i]_m = f([Q(x_a)]_m) = s_m f(x_{a,\sigma(m)}) = [A_Q b_a]_m,
    \]
    so $\tilde{b}_i = A_Q b_a$.
    \item \emph{Bounded residual update and clipping:} Since $A_Q$ is a signed permutation matrix and $\tanh$ is an odd coordinate-wise function ($\tanh(-z) = -\tanh(z)$), we have $\tanh(A_Q y) = A_Q \tanh(y)$ for all $y \in \R^d$. Hence the pre-clip update satisfies
    \[
    \tilde{x}_i + s\tanh(\tilde{v}_i + \gamma_\ell \tilde{b}_i) = A_Q x_a + c_Q + s A_Q \tanh(v_a + \gamma_\ell b_a) = Q\!\big(x_a + s\tanh(v_a + \gamma_\ell b_a)\big).
    \]
    Finally, for any $y \in \R$, $\clip(1 - y) = 1 - \clip(y)$, so $\clip(Q(z)) = Q(\clip(z))$ coordinate-wise. Thus $\tilde{x}_i^{(\ell)} = Q(x_{\pi^{-1}(i)}^{(\ell)})$, proving layer-wise $(S_n \times B_d)$-equivariance.
\end{enumerate}
By induction over $\ell = 1,\dots,L$, $G_\theta$ is $(S_n \times B_d)$-equivariant. For $\Restore(Z, X^{(0)})$ in \eqref{eq:restore}, because $Q$ is an affine isometry, $(1-\alpha)Q(z_a) + \alpha Q(x_a^{(0)}) = Q((1-\alpha)z_a + \alpha x_a^{(0)})$, and exact row equality (i.e., $\delta = 0$) is $(S_n \times B_d)$-invariant; hence the selected $\alpha^*$ is identical and $\Restore(P\cdot Q(Z), P\cdot Q(X^{(0)})) = P\cdot Q(\Restore(Z, X^{(0)}))$. The same affine identity applies to the residual gate $X + \tanh(g)(H_\theta(X) - X)$ followed by $\clip$ and $\Restore$, establishing exact $(S_n \times B_d)$-equivariance of $H_\theta$ and $R_\theta$.

\textbf{Part (iii): Parameter count.}
Each of the $L$ layers contains \texttt{Linear(5, W)} ($5W + W = 6W$ parameters), \texttt{Linear(W, 1)} ($W + 1$ parameters), and one scalar $\gamma_\ell$ ($1$ parameter), totaling $7W + 2$ parameters per layer and $|\theta_G| = L(7W + 2)$ overall. Adding the single scalar gate $g$ in $R_\theta$ gives $|\theta_R| = L(7W + 2) + 1$. Neither expression depends on $n$ or $d$. For instance, $(L,W)=(5,32) \Rightarrow 1{,}130$; $(10,32) \Rightarrow 2{,}260$ ($2{,}261$ with gate); $(16,128) \Rightarrow 14{,}368$ ($14{,}369$ with gate).
\end{proof}

\subsection{Proof of Theorem~\ref{thm:feasibility} (Finite-Step Feasibility and Zero-Bias Barrier)}
\begin{proof}
\textbf{Part (i):} Since $\cube$ is convex and $Z, X^{(0)} \in \bigl(\cube\bigr)^n$, every convex combination $Z(\alpha) = (1-\alpha)Z + \alpha X^{(0)}$ with $\alpha \in [0,1]$ lies in $\bigl(\cube\bigr)^n$. Because $\alpha = 1 \in \mathcal{A}$ yields $Z(1) = X^{(0)}$ and $\delta(X^{(0)}) > 0$, the predicate $\delta(Z(\alpha)) > 0$ holds at least at $\alpha = 1$, guaranteeing termination in at most $|\mathcal{A}| = 6$ trials with $\delta(\Restore(Z, X^{(0)})) > 0$.

Furthermore, for any distinct pair $i \neq j$, consider the vector difference $d_{ij}(\alpha) = (1-\alpha)(z_i - z_j) + \alpha(x_i^{(0)} - x_j^{(0)}) = (z_i - z_j) + \alpha \big((x_i^{(0)} - x_j^{(0)}) - (z_i - z_j)\big)$. Since $x_i^{(0)} \neq x_j^{(0)}$:
\begin{itemize}
    \item If $z_i - z_j = x_i^{(0)} - x_j^{(0)}$, then $d_{ij}(\alpha) = x_i^{(0)} - x_j^{(0)} \neq 0$ for all $\alpha \in [0,1]$.
    \item Otherwise, $w_{ij} = (x_i^{(0)} - x_j^{(0)}) - (z_i - z_j) \neq 0$, so the affine line $\alpha \mapsto (z_i - z_j) + \alpha w_{ij}$ passes through $0 \in \R^d$ for at most one value of $\alpha \in \R$.
\end{itemize}
Summing over all $\binom{n}{2}$ unordered pairs $(i,j)$, at most $\binom{n}{2}$ values of $\alpha \in [0,1)$ can induce any collision.

\textbf{Part (ii):} For $r > 0$, $\tilde{\psi}_{\mathrm{log}}(r) = (\log(r/\tau))^2$ for $r \in (0, \tau)$ and $\tilde{\psi}_{\mathrm{log}}(r) = 0$ for $r \ge \tau$. On $(0, \tau)$, it is smooth with derivative $\tilde{\psi}'_{\mathrm{log}}(r) = \frac{2}{r}\log(r/\tau) < 0$. As $r \to \tau^-$, both $\tilde{\psi}_{\mathrm{log}}(r) \to (\log 1)^2 = 0$ and $\tilde{\psi}'_{\mathrm{log}}(r) \to \frac{2}{\tau}\log 1 = 0$, matching the constant-zero branch on $[\tau, \infty)$. Thus $\tilde{\psi}_{\mathrm{log}} \in C^1((0,\infty))$, diverges as $(\log r)^2 \to +\infty$ when $r \to 0^+$, and vanishes identically in value and gradient for all $r \ge \tau$.
\end{proof}

\subsection{Proof of Theorem~\ref{thm:shared_witness} (Shared-Witness Paired Surrogate)}
\begin{proof}
\textbf{Part (i):} Since $U = \mathcal{S}(X) \cup \mathcal{S}(Z) \subset \cube$, taking the maximum over a finite subset of $\cube$ yields $\hat{h}_U(Y) = \max_{u \in U}\min_i \|u - y_i\|_2 \le \sup_{u \in \cube}\min_i \|u - y_i\|_2 = h(Y)$ for any $Y$. Moreover, because $\mathcal{S}(X) \subseteq U$ and $\mathcal{S}(Z) \subseteq U$, monotonicity of the maximum over sets gives $\hat{h}_U(Z) = \max\{\hat{h}_{\mathcal{S}(X)}(Z), \hat{h}_{\mathcal{S}(Z)}(Z)\}$.

\textbf{Part (ii):} When $Z = X$, $\hat{h}_U^2(Z) - \hat{h}_U^2(X) = \hat{h}_U^2(X) - \hat{h}_U^2(X) = 0$ pointwise for every realization of $U$, so the degradation hinge has zero value and zero variance. Conversely, if $U_1, U_2$ are i.i.d.\ random witness sets with non-degenerate empirical fill $\hat{h}_{U_1}^2(X) \stackrel{d}{=} \hat{h}_{U_2}^2(X)$ and $\operatorname{Var}(\hat{h}_{U_1}^2(X)) > 0$, then $D = \hat{h}_{U_2}^2(X) - \hat{h}_{U_1}^2(X)$ is symmetric around $0$ with $\Pr(D > 0) > 0$, whence $\E[[D]_+] = \frac{1}{2}\E[|D|] > 0$.

\textbf{Part (iii):} Treating $U$ and $\hat{h}_U^2(X)$ as constants with respect to $\theta$ (since $\mathcal{S}$ runs under \texttt{no\_grad}), the coverage portion of $\mathcal{L}_{\mathrm{inc}}$ is $f(Z) = \hat{h}_U^2(Z) + \lambda_{\mathrm{reg}}[\hat{h}_U^2(Z) - \hat{h}_U^2(X)]_+$. Whenever $\hat{h}_U^2(Z) > \hat{h}_U^2(X)$, $f(Z) = (1 + \lambda_{\mathrm{reg}})\hat{h}_U^2(Z) - \lambda_{\mathrm{reg}}\hat{h}_U^2(X)$, so by Danskin's theorem any Clarke subgradient satisfies
\[
\partial_Z f(Z) = (1 + \lambda_{\mathrm{reg}})\,\partial_Z \hat{h}_U^2(Z),
\]
amplifying the active worst-witness gradient by $1 + \lambda_{\mathrm{reg}} = 5$ relative to non-degrading designs ($\hat{h}_U^2(Z) < \hat{h}_U^2(X)$).
\end{proof}

\subsection{Proof of Proposition~\ref{prop:multiloop} (Multi-Loop Contraction and Guarded Bound)}
\begin{proof}
\textbf{Part (i):} Unrolling $e_k := h(X^{(k)}) - h^* \le \gamma e_{k-1} + \eta$ recursively for $k = 1, \dots, K$ yields
\[
e_K \le \gamma^K e_0 + \eta \sum_{j=0}^{K-1} \gamma^j = \gamma^K e_0 + \frac{1 - \gamma^K}{1 - \gamma}\eta.
\]
\textbf{Part (ii):} Let $k^* \in \arg\min_{0 \le k \le K} h(X^{(k)})$ be an optimal iterate along the trajectory and $\hat{k} = \arg\min_{0 \le k \le K} \hat{h}_k$ be the proxy-selected iterate. Using the assumed lower-bound error $h(X^{(k)}) - \varepsilon_k \le \hat{h}_k \le h(X^{(k)})$:
\[
h(X^{(\hat{k})}) - \varepsilon_{\hat{k}} \;\le\; \hat{h}_{\hat{k}} \;\le\; \hat{h}_{k^*} \;\le\; h(X^{(k^*)}) = \min_{0 \le k \le K} h(X^{(k)}),
\]
which rearranges directly to $h(X^{(\hat{k})}) \le \min_{0 \le k \le K} h(X^{(k)}) + \varepsilon_{\hat{k}}$. By induction on $k$ using Theorem~\ref{thm:feasibility}(i), every iterate $X^{(k)} = R_\theta(X^{(k-1)})$ has $\delta(X^{(k)}) > 0$ whenever $\delta(X^{(0)}) > 0$, so $\delta(X^{(\hat{k})}) > 0$. In $d=2$ with exact Voronoi scoring, $\hat{h}_k = h(X^{(k)})$ so $\varepsilon_k = 0$ identically.
\end{proof}

\section{Detailed Architecture Specifications and Training Protocols}
\label{app:archs}

Table~\ref{tab:arch_summary} summarizes every architecture evaluated across the fixed benchmark (V1--V7), controlled mixing ablations, and 10-hour compute-matched budget sweep.

\begin{table}[h]
\centering
\caption{Summary of all neural architectures evaluated in this study.}
\label{tab:arch_summary}
\small
\begin{tabular}{@{}llrrl@{}}
\toprule
Model & Family & Layers & Params & Inference Interface / Key Mechanism \\
\midrule
V1 & Bipartite GNN & 1 & 1,283 & Joint $X$/$U$ complete bipartite graph; 256 probes + probe head \\
V2 & Context-Step GNN & 2 & 4,710 & V1 + node context, secondary aggregation, per-node step gate \\
V3 & Geom.\ Transformer & 2 & 147,676 & Width 64, 4 heads; geometric-biased self/cross attention ($X,U$) \\
V4 & Large Transformer & 6 & 6,530,964 & Width 256, 8 heads; BF16 + activation checkpointing ($X,U$) \\
V5 & Lattice-Teacher V4 & 6 & 6,530,964 & V4 architecture trained with interleaved lattice teachers + noise \\
V6 & Sparse-Edge GNN & 4 & 9,090 & 2-NN blue--blue + 2-NN blue--red MLPs; numerical probe ascent \\
\midrule
V7 Base ($5\times 32$) & Equiv.\ $X$-Only GNN & 5 & 1,130 & Pure $X\to X$, $(S_n\times B_d)$-equiv., witness search train-only \\
V7 $10\times 32$ (+ gate) & Equiv.\ $X$-Only GNN & 10 & 2,261 & Depth-scaled V7 + Hard $\Restore$ + residual gate $g$ \\
V7 $16\times 128$ (+ gate) & Equiv.\ $X$-Only GNN & 16 & 14,369 & BO-selected V7 + Hard $\Restore$ + residual gate $g$ \\
\midrule
\texttt{TransformerRefiner} & Dim.-Agnostic Transf. & 6 & 1,239,553 & Coord.\ MLP $(x,\min(x,1-x))$, mean-pooled 4-head Transf., Hard $\Restore$ \\
\texttt{GeomTransformer v2} & Geom.-Biased Transf. & 6 & 1,211,574 & Coord.\ MLP encoder + V7 5D attn.\ bias + rel.\ updates, Hard $\Restore$ \\
\bottomrule
\end{tabular}
\end{table}

\paragraph{Training Hyperparameters.}
\begin{itemize}
    \item \emph{Fixed Benchmark (V1--V7, Table~\ref{tab:fixed_sep}):} 8 epochs over 1,160 training specs (batch size 2 per spec, fixed inputs), AdamW $\mathrm{lr}=3\times 10^{-4}$ ($\mathrm{lr}=10^{-4}$ for separation/Hard continuations), gradient clip 2.0. Witness search for V7: $M_{\mathrm{pool}}=512$ candidates (mixed Sobol'/uniform, faces, corners), $M_w=48$ starts, $T_w=8$ steps with step size $0.32\,s\cdot 0.78^t$. Expanding the witness search $4\times$ to $(M_{\mathrm{pool}}=2048, M_w=192)$ in a controlled Epoch-10 paired trial yielded no net gain on the 9 held-out Mak specs ($+0.279\%$ fill, 3/9 wins), confirming that $(512, 48)$ suffices.
    \item \emph{Streaming V7-Mak Incremental Refinement (Section~\ref{sec:exp_inc}):} Warm-started from Epoch-18 $10\times 32$ Hard V7 (`lr=1e-4`), $g_0 = 0$, 14h wall clock (397,812 single-layout updates, peak GPU reserved memory $0.42\,\mathrm{GiB}$) with 6 concurrent CPU workers generating Mak+PSO layouts.
    \item \emph{10-Hour Mixed Budget Sweep (Section~\ref{sec:exp_mix}):} 50\% uniform random batches paired with 50\% structured batches cycling Mak+PSO (1,352 odd-$n$ layouts), He 2017 lattices (77 odd-$n$ layouts, $d\in[2,5]$), and fresh scrambled Sobol' designs. Adaptive edge-budget batching ($B = \operatorname{clamp}(\lfloor E/n^2\rfloor, 1, 16384)$) dynamically adjusts $E$ to maintain $\approx 84\%$ GPU memory utilization. AdamW $\mathrm{lr}=10^{-4}$, FP64 global gradient norm clipping at 2.0, $\lambda_{\mathrm{reg}}=4$, $\lambda_{\mathrm{sep}}=0.02$, $\tau=0.5$.
\end{itemize}

\section{Complete Per-Specification Held-Out Results and Classical Method Comparisons}
\label{app:per_spec}

Table~\ref{tab:per_spec_9} reports per-specification fill distances on all nine held-out even specifications for classical paper methods, our 14h incremental V7-Mak refiner, and our 10h mixed V7 $16\times 128$ operator under single-pass ($K=1$) and multi-loop ($K=2$, PG-16) inference from pure uniform random noise. Table~\ref{tab:t56_sep} documents the simultaneous fill-distance reduction and separation-distance increase achieved by incremental V7-Mak on these nine specs.

\begin{table}[h]
\centering
\caption{\textbf{Per-specification held-out comparison} across all 9 even specifications ($d \in \{2,3,5,8\}$). Bold marks the lowest fill distance among arbitrary-$(n,d)$ methods (excluding He's algebraic lattices, which are shown separately as structural reference where available and require lattice-compatible sizes $n' \le n$).}
\label{tab:per_spec_9}
\setlength{\tabcolsep}{4.2pt}
\footnotesize
\begin{tabular}{@{}lrrrrrrrrr@{}}
\toprule
$(d,n)$ & Rand. & FSSF & Mak & Mak & SFDes. & V7-Mak & V7 $16\times 128$ & V7 $16\times 128$ & He 2017 \\
 & ($K=0$) & \citep{shang2021fssf} & L-BFGS & +PSO & \citep{wang2026sfdesign} & (Incr.) & (Rand, $K{=}2$) & (Rand, PG-16) & (Lattice*) \\
\midrule
$(2,30)$ & 0.3705 & 0.2012 & 0.1435 & \textbf{0.1384} & 0.1433 & 0.1397 & 0.1422 & 0.1407 & 0.1266 \\
$(2,40)$ & 0.2915 & 0.1782 & 0.1267 & 0.1232 & 0.1255 & 0.1210 & \textbf{0.1201} & 0.1206 & 0.1094$^\dagger$ \\
$(2,50)$ & 0.2745 & 0.1428 & 0.1082 & 0.1102 & 0.1110 & 0.1080 & 0.1095 & \textbf{0.1049} & 0.0963$^\dagger$ \\
$(2,100)$ & 0.1562 & 0.0961 & 0.0840 & 0.0825 & 0.0902 & 0.0790 & 0.0768 & \textbf{0.0759} & 0.0659 \\
$(3,30)$ & 0.5152 & 0.3212 & 0.3146 & 0.3143 & 0.3276 & \textbf{0.3028} & 0.3131 & 0.3032 & 0.2795 \\
$(5,100)$ & 0.7775 & 0.5726 & 0.5595 & 0.5512 & 0.5826 & 0.5156 & 0.5127 & \textbf{0.5101} & 0.4656$^\dagger$ \\
$(8,30)$ & 1.3635 & 1.2783 & 1.2086 & 1.2159 & \textbf{1.2007} & 1.2126 & 1.2504 & 1.2248 & 1.0000$^\dagger$ \\
$(8,50)$ & 1.2496 & 1.1839 & 1.1400 & 1.1515 & 1.1466 & \textbf{1.1250} & 1.1858 & 1.1477 & 0.9014$^\dagger$ \\
$(8,100)$ & 1.1632 & 1.0836 & 1.0665 & 1.0402 & 1.0708 & 0.9978 & \textbf{0.9744} & 0.9766 & 0.8496$^\dagger$ \\
\bottomrule
\end{tabular}
\par\vspace{2pt}\raggedright\scriptsize
$^*$He (2017) \texttt{InterleavedMinimaxD} is restricted to $d \le 8$; $^\dagger$denotes padded designs where the author package returned fewer than $n$ points ($n' \in \{39,49,90,24,48,96\}$).
\end{table}

\begin{table}[h]
\centering
\caption{\textbf{Simultaneous fill-distance reduction and separation-distance improvement} of 14h incremental V7-Mak on the 9 held-out Mak+PSO specifications (3 replicates per spec).}
\label{tab:t56_sep}
\small
\begin{tabular}{@{}lrrrrrr@{}}
\toprule
$(d,n)$ & Mak $h$ & Refined $h$ & Fill Change & Mak $\delta$ & Refined $\delta$ & Sep.\ Change \\
\midrule
$(2,30)$ & 0.13844 & 0.13971 & $+0.95\%$ & 0.15008 & 0.15414 & $+2.7\%$ \\
$(2,40)$ & 0.12315 & 0.12098 & $-1.67\%$ & 0.11212 & 0.11413 & $+1.8\%$ \\
$(2,50)$ & 0.11022 & 0.10801 & $-2.01\%$ & 0.10519 & 0.10643 & $+1.2\%$ \\
$(2,100)$ & 0.08252 & 0.07904 & $-4.21\%$ & 0.06088 & 0.06034 & $-0.9\%$ \\
$(3,30)$ & 0.31432 & 0.30278 & $-3.58\%$ & 0.24505 & 0.27222 & $+11.1\%$ \\
$(5,100)$ & 0.55123 & 0.51555 & $-6.43\%$ & 0.28140 & 0.33425 & $+18.8\%$ \\
$(8,30)$ & 1.21593 & 1.21261 & $-0.27\%$ & 0.60454 & 0.78007 & $+29.0\%$ \\
$(8,50)$ & 1.15148 & 1.12496 & $-2.30\%$ & 0.55170 & 0.72065 & $+30.6\%$ \\
$(8,100)$ & 1.04019 & 0.99778 & $-4.07\%$ & 0.46825 & 0.56033 & $+19.7\%$ \\
\bottomrule
\end{tabular}
\end{table}

\section{Architecture Scaling, 19-Trial BO, and Arm 0 Multi-Loop Trajectory}
\label{app:bo}

\paragraph{19-Trial Bayesian Architecture Search.}
We executed a sequential Mat\'ern-5/2 GP Expected Improvement search over 16 V7 configurations $(L \in \{5,10,16,24\}, W \in \{32,64,128,192\})$ and 20 Transformer configurations $(L \in \{2,4,6,8\}, W \in \{32,64,96,128,192\})$, training each trial from scratch for 30 minutes under mixed structured/random batches and evaluating on 48 held-out odd validation specifications (equally weighting Mak-start and Random-start reductions vs.\ Mak+PSO). Table~\ref{tab:bo19} lists all 19 completed trials: \textbf{V7 $16\times 128$} achieved the highest validation objective ($-3.82\%$), outperforming all 4 Transformers by more than 12 percentage points. Very large V7 models ($16\times 192$, $24\times 192$) underperformed under the 30-minute wall-clock cap due to lower layout throughput.

\begin{table}[h]
\centering
\caption{\textbf{All 19 trials of Bayesian architecture optimization} (30 min from scratch per trial, evaluated on 48 held-out validation specs). Positive values indicate improvement over Mak+PSO.}
\label{tab:bo19}
\setlength{\tabcolsep}{4pt}
\footnotesize
\begin{tabular}{@{}rlrrrrrr@{}}
\toprule
Trial & Architecture & Params & Steps & Mak vs.\ Mak & Rand vs.\ Mak & Joint Obj. & Wins (Mak/Rand) \\
\midrule
\textbf{2} & \textbf{V7 $16\times 128$} & \textbf{14,369} & \textbf{7,800} & \textbf{$+5.51\%$} & \textbf{$-13.15\%$} & \textbf{$-3.82\%$} & \textbf{38 / 13} \\
10 & V7 $10\times 192$ & 13,461 & 7,608 & $+5.16\%$ & $-15.08\%$ & $-4.96\%$ & 35 / 11 \\
8 & V7 $10\times 128$ & 8,981 & 7,066 & $+5.34\%$ & $-15.64\%$ & $-5.15\%$ & 36 / 10 \\
11 & V7 $5\times 192$ & 6,731 & 6,722 & $+4.68\%$ & $-15.03\%$ & $-5.17\%$ & 36 / 10 \\
9 & V7 $16\times 64$ & 7,201 & 5,717 & $+5.16\%$ & $-17.39\%$ & $-6.12\%$ & 35 / 11 \\
17 & V7 $24\times 64$ & 10,801 & 6,072 & $+5.94\%$ & $-18.36\%$ & $-6.21\%$ & 40 / 8 \\
16 & V7 $10\times 64$ & 4,501 & 5,293 & $+4.85\%$ & $-18.07\%$ & $-6.61\%$ & 36 / 8 \\
4 & V7 $5\times 64$ & 2,251 & 4,867 & $+5.09\%$ & $-18.58\%$ & $-6.74\%$ & 36 / 8 \\
14 & V7 $5\times 128$ & 4,491 & 6,215 & $+4.07\%$ & $-17.60\%$ & $-6.76\%$ & 34 / 11 \\
12 & V7 $24\times 32$ & 5,425 & 5,142 & $+5.57\%$ & $-20.14\%$ & $-7.29\%$ & 40 / 6 \\
13 & V7 $5\times 32$ & 1,131 & 3,738 & $+5.20\%$ & $-21.29\%$ & $-8.04\%$ & 39 / 5 \\
1 & V7 $10\times 32$ & 2,261 & 4,392 & $+5.55\%$ & $-22.31\%$ & $-8.38\%$ & 37 / 6 \\
19 & V7 $16\times 32$ & 3,617 & 4,814 & $+3.27\%$ & $-25.09\%$ & $-10.91\%$ & 32 / 1 \\
7 & V7 $24\times 192$ & 32,305 & 6,772 & $-2.61\%$ & $-21.88\%$ & $-12.25\%$ & 20 / 2 \\
18 & V7 $16\times 192$ & 21,537 & 7,608 & $-5.40\%$ & $-26.15\%$ & $-15.77\%$ & 20 / 0 \\
\midrule
6 & Transf.\ $4\times 192$ & 1,891,201 & 5,469 & $+2.85\%$ & $-34.53\%$ & $-15.84\%$ & 27 / 0 \\
3 & Transf.\ $6\times 128$ & 1,239,553 & 5,267 & $+4.32\%$ & $-36.11\%$ & $-15.89\%$ & 40 / 0 \\
15 & Transf.\ $8\times 32$ & 104,897 & 2,622 & $+4.23\%$ & $-37.34\%$ & $-16.55\%$ & 41 / 0 \\
5 & Transf.\ $2\times 64$ & 112,641 & 2,511 & $+4.02\%$ & $-37.44\%$ & $-16.71\%$ & 43 / 0 \\
\bottomrule
\end{tabular}
\end{table}

\paragraph{Six-Architecture Fixed-Benchmark Sweep (Epoch 36).}
\label{app:sweep}
Starting from Epoch-12 $10\times 32$ Hard V7 on random-only fixed-benchmark inputs, we compared six extensions through Epoch 36 on the 9 held-out Mak specs: $10\times 32$ control ($+4.77\%$ above Mak+PSO, 2/9 wins), $16\times 32$ depth ($+5.42\%$, 3/9), $10\times 128$ width ($+11.51\%$, 3/9), $16\times 128$ depth+width ($+6.37\%$, 3/9), $16\times 128$ + residual edge MLP ($+13.62\%$, 1/9), and $16\times 128$ + 2-layer Transformer head ($+4.65\%$, 2/9). All six beat FSSF on 8/9 specs ($-6.9\%$ to $-8.7\%$) at Epoch 33, but remained $4.7\%$--$13.6\%$ behind Mak+PSO without incremental anchor training and random mixing—confirming that Pillars~2 and~3, not raw parameter scaling alone, drive the breakthrough over Mak+PSO.

\paragraph{Arm 0 Structured-Only 100-Loop Trajectory.}
\label{app:multiloop_arm0}
Table~\ref{tab:arm0_100} shows open-loop recursion up to $K=100$ for Arm~0 (V7 $10\times 32$ trained 24h on Mak+He anchors \emph{without} random inputs, 874,420 steps). Multi-loop self-feedback reduces its held-out random-input fill distance from $-12.71\%$ ($K=1$) to $-20.76\%$ ($K=4$) and slashes its gap to Mak+PSO on seen specs from $+17.56\%$ down to $+2.86\%$ (38--42/96 wins), before gently drifting to $-16.54\%$ at $K=100$—directly illustrating Proposition~\ref{prop:multiloop}(i).

\begin{table}[h]
\centering
\caption{\textbf{Open-loop 100-iteration trajectory of Arm 0} (structured-only V7 $10\times 32$, 874K steps) on 96 held-out and 96 seen random-input specifications (288 layouts each).}
\label{tab:arm0_100}
\small
\begin{tabular}{@{}lrrrrrrrr@{}}
\toprule
Metric & $K=1$ & $K=2$ & $K=4$ & $K=8$ & $K=16$ & $K=32$ & $K=64$ & $K=100$ \\
\midrule
Held-out vs.\ Rand.\ (\%) & $-12.71$ & $-19.31$ & $\mathbf{-20.76}$ & $-18.60$ & $-17.01$ & $-16.61$ & $-16.49$ & $-16.54$ \\
Seen vs.\ Rand.\ (\%) & $-12.94$ & $-19.66$ & $\mathbf{-21.01}$ & $-18.24$ & $-16.95$ & $-16.58$ & $-16.39$ & $-16.33$ \\
Seen gap vs.\ Mak (\%) & $+17.56$ & $+6.98$ & $\mathbf{+2.86}$ & $+5.10$ & $+6.28$ & $+6.59$ & $+6.75$ & $+6.77$ \\
Seen wins vs.\ Mak (/96) & 16 & 40 & 38 & 40 & 40 & 41 & 42 & 42 \\
\bottomrule
\end{tabular}
\end{table}

\section{Zero-Shot Unseen Dimension and Large-$n$ Breakdown}
\label{app:ood}

Table~\ref{tab:ood_full} provides exact fill distances across all 30 unseen-dimension specs ($d \in \{6,11,12\}$) and 5 large-$n$ extrapolation specs ($n \in \{500, 1000\}$) summarized in Section~\ref{sec:exp_ood}.

\begin{table}[h]
\centering
\caption{\textbf{Zero-shot unseen-dimension ($d\in\{6,11,12\}$, 10 budgets each) and large-$n$ extrapolation} of 10-hour V7 models vs.\ classical and Transformer baselines (mean fill distance $h$, lower is better).}
\label{tab:ood_full}
\small
\begin{tabular}{@{}lrrrrrrrr@{}}
\toprule
Regime & Random & LHS & Sobol' & FF & Transf. & Geom-T v2 & V7 $10\times 32$ & V7 $16\times 128$ \\
 & & & & & ($K=1$) & ($K=1$) & ($K=1$ / $K=2$) & ($K=1$ / $K=2$) \\
\midrule
Unseen $d=6$ ($n=10\dots 300$) & 0.9736 & 0.9237 & 0.9078 & 0.8380 & 0.9140 & 0.8984 & 0.7556 / 0.7561 & 0.7398 / \textbf{0.7328} \\
Unseen $d=11$ ($n=10\dots 300$) & 1.5661 & 1.5305 & 1.5117 & 1.5272 & 1.5184 & 1.5390 & 1.4583 / 1.4407 & 1.4487 / \textbf{1.4321} \\
Unseen $d=12$ ($n=10\dots 300$) & 1.6520 & 1.6356 & 1.6063 & 1.6340 & 1.6057 & 1.6316 & \textbf{1.5347} / 1.6245 & 1.5362 / 1.5393 \\
\midrule
$(d=2, n=500)$ & 0.0894 & --- & 0.0562 & --- & --- & --- & 0.0371 / \textbf{0.0340} & 0.0480 / 0.0451 \\
$(d=5, n=500)$ & 0.5210 & --- & 0.4688 & --- & --- & --- & \textbf{0.3280} / 0.3442 & 0.3448 / 0.3535 \\
$(d=8, n=500)$ & 0.9573 & --- & 0.9026 & --- & --- & --- & 0.7046 / \textbf{0.6955} & 0.7433 / 0.7183 \\
$(d=2, n=1000)$ & 0.0590 & --- & 0.0382 & --- & --- & --- & 0.0399 / 0.0370 & 0.0481 / \textbf{0.0365} \\
$(d=5, n=1000)$ & 0.4505 & --- & 0.4265 & --- & --- & --- & 0.3089 / 0.3017 & 0.3087 / \textbf{0.2871} \\
\bottomrule
\end{tabular}
\end{table}

\end{document}